\documentstyle[preprint,prb,aps,eqsecnum,amsmath,amssymb,epsfig,psfrag,overcite]{revtex}

\tighten

\draft

\newcommand{\comment}[1]{}

\newcommand{\eqlabel}[1]{\label{eq:#1}}

\newcommand{\seclabel}[1]{\label{sec:#1}}
\newcommand{\figlabel}[1]{\label{fig:#1}}

\renewcommand{\eqref}[1]{(\ref{eq:#1})}

\newcommand{\secref}[1]{Sec.~\ref{sec:#1}}
\newcommand{\secreftwo}[2]{Secs.~\ref{sec:#1} and~\ref{sec:#2}}

\newcommand{\figref}[1]{Fig.~\ref{fig:#1}}
\newcommand{\figreftwo}[2]{Figs.~\ref{fig:#1} and~\ref{fig:#2}}

\newcommand{\etal}{\textit{et~al.}}
\newcommand{\ie}{\textit{i.e.}}

\newcommand{\mathnotation}[2]{\newcommand{#1}{#2}}

\renewcommand{\onlinecite}[1]{\citen{#1}}

\renewcommand{\l}{\left}			
\renewcommand{\r}{\right}			
\mathnotation{\pd}{\partial}			
\mathnotation{\ee}{{\mathrm e}}			
\mathnotation{\grad}{{\boldsymbol{\nabla}}}		
\mathnotation{\ldef}{\mathrel{\raisebox{.069ex}{:}\!\!=}}
\mathnotation{\rdef}{\mathrel{=\!\!\raisebox{.069ex}{:}}}
\mathnotation{\half}{{\textstyle {1\over2}}}	
\mathnotation{\levicivita}{\varepsilon}		
\mathnotation{\curl}{\grad\times}		
\renewcommand{\div}{\grad\cdot}			
\mathnotation{\pheq}{&\phantom{=}}		
\renewcommand{\time}{t}				
\mathnotation{\Diff}{D}				
\mathnotation{\lyapexp}{\lambda}		
\mathnotation{\lyapexpt}{\eta}
\mathnotation{\lyapexpinfty}{\lyapexp^\infty}	
\mathnotation{\Riccigr}{\Delta}			
\mathnotation{\Kdivs}{K}
\mathnotation{\x}{x}				
\mathnotation{\xv}{{\mathbf{\x}}}		
\mathnotation{\vel}{{\mathbf{v}}}		
\mathnotation{\velc}{v}				
\mathnotation{\velocr}{\vel_{\mathrm{ocr}}}	
\mathnotation{\sdim}{n}				
\mathnotation{\w}{e}				
\mathnotation{\wu}{\hat{\w}}			
\mathnotation{\wv}{{\mathbf{\w}}}
\mathnotation{\wuv}{{\mathbf{\wu}}}
\mathnotation{\V}{V}				
\mathnotation{\lagrc}{\xi}			
\mathnotation{\lagrcv}{{\boldsymbol{\lagrc}}}	
\mathnotation{\metric}{g}			
\mathnotation{\detmetric}{g}			
\mathnotation{\flatmetric}{\delta}		
\mathnotation{\geigen}{\Lambda}			
\mathnotation{\sdir}{s}				
\mathnotation{\edir}{u}				
\mathnotation{\mdir}{m}				
\mathnotation{\sdirv}{{\mathbf{\sdir}}}
\mathnotation{\edirv}{{\mathbf{\edir}}}
\mathnotation{\mdirv}{{\mathbf{\mdir}}}
\mathnotation{\sdiru}{\hat{\sdir}}
\mathnotation{\ediru}{\hat{\edir}}
\mathnotation{\mdiru}{\hat{\mdir}}
\mathnotation{\sdiruv}{{\mathbf{\sdiru}}}
\mathnotation{\ediruv}{{\mathbf{\ediru}}}
\mathnotation{\mdiruv}{{\mathbf{\mdiru}}}
\mathnotation{\Rcurv}{R}			
\mathnotation{\Ricci}{R}			
\mathnotation{\Christoffel}{\Gamma}		
\mathnotation{\Riccirotc}{\omega}		
\mathnotation{\gradlc}{\grad}			
\mathnotation{\curllc}{\gradlc\times}		
\mathnotation{\divlc}{\gradlc\cdot}		
\mathnotation{\covder}{\nabla}			
\mathnotation{\kcurvm}{\kappa}			
\mathnotation{\kcurv}{{\boldsymbol{\kcurvm}}}	
\mathnotation{\helic}{{\mathcal{H}}}		
\mathnotation{\helict}{\widetilde{\helic}}

\title{Geometrical Constraints on Finite-time Lyapunov Exponents in Two and
Three Dimensions}

\author{Jean-Luc Thiffeault and Allen H. Boozer}

\address{Columbia University, Department of Applied Physics and Applied
Mathematics,\\ New York, NY 10027}

\begin{document}

\maketitle

\begin{abstract}

Constraints are found on the spatial variation of finite-time Lyapunov
exponents of two- and three-dimensional systems of ordinary differential
equations.  In a chaotic system, finite-time Lyapunov exponents describe the
average rate of separation, along characteristic directions, of neighboring
trajectories.  The solution of the equations is a coordinate transformation
that takes initial conditions (the Lagrangian coordinates) to the state of the
system at a later time (the Eulerian coordinates).  This coordinate
transformation naturally defines a metric tensor, from which the Lyapunov
exponents and characteristic directions are obtained.  By requiring that the
Riemann curvature tensor vanish for the metric tensor (a basic result of
differential geometry in a flat space), differential constraints relating the
finite-time Lyapunov exponents to the characteristic directions are derived.
These constraints are realized with exponential accuracy in time.  A
consequence of the relations is that the finite-time Lyapunov exponents are
locally small in regions where the curvature of the stable manifold is large,
which has implications for the efficiency of chaotic mixing in the
advection--diffusion equation.  The constraints also modify previous estimates
of the asymptotic growth rates of quantities in the dynamo problem, such as
the magnitude of the induced current.

\end{abstract}

\vspace{.5em}

\begin{center}
\textit{Published in \emph{Chaos~\textbf{11}, 16-28 (2001)}.
Copyright 2001 American Institute of Physics}
\end{center}


\vspace{.5em}

{\bf In a dynamical system, the average rate of stretching of infinitesimal
elements of phase-space is described by the Lyapunov exponents.  For a fluid,
this stretching is crucial to the enhanced diffusion observed in chaotic
flows.  Though the exponents converge to a time-asymptotic limit that is
independent of the initial condition, in most situations of interest this
convergence is very slow: there exist regions where, for finite times, the
exponents remain anomalously small.  Information on the spatial and temporal
behavior of finite-time Lyapunov exponents is thus useful in identifying such
regions of low stretching rates.  Our approach is to regard the time evolution
of the solution to a set of ordinary differential equations as a coordinate
transformation.  For smooth dynamical systems, we can study this
transformation with the tools of differential geometry.  Specifically, we
consider the Riemann curvature tensor, a quantity that must vanish in every
coordinate system when using the usual Euclidean distance.  By examining the
details of how the different terms of the Riemann tensor balance each other,
we obtain constraints on the finite-time Lyapunov exponents, clarifying and
generalizing previous work.\cite{Tang1996,Tang1999b} These constraints take
the form of differential relations between the exponents and the
characteristic directions of stretching in the flow.  The constraints have
implications for chaotic mixing, where they allow identification of locally
small finite-time Lyapunov exponents, and for the kinematic dynamo problem,
where they imply a slower growth rate of the power needed to sustain a dynamo.
}

\section{Introduction}

Lyapunov exponents are fundamental to our understanding of chaotic processes.
They describe the time-asymptotic rate of separation of neighboring
trajectories in a dynamical system, and are global properties of a chaotic
region, independent of the trajectory chosen to measure them (the fiducial
trajectory).  This independence is a consequence of a theorem of
Oseledec,\cite{Oseledec1968} which applies in the limit of infinite time.  In
contrast, the \emph{finite-time} Lyapunov exponents, which describe the
instantaneous rate of separation of neighboring trajectories, depend on the
choice of fiducial trajectory and so characterize local properties of the
chaotic region.  Similarly, the characteristic direction of expansion or
contraction associated with each exponent is a local property of the chaotic
region; it converges exponentially to its time-asymptotic
orientation\cite{Greene1987,Goldhirsch1987}---a much more rapid convergence
rate than that of the Lyapunov exponents.

The slow convergence of the Lyapunov exponents means that in practical
situations, \ie, a computer or laboratory experiment, the system under study
only rarely evolves long enough to ``feel'' the value of the true,
infinite-time Lyapunov exponents.  Rather, the rates of stretching and
contracting of infinitesimal elements of phase-space are dominated by local
effects, described by the finite-time Lyapunov exponents.  In fluid dynamics
and plasma physics, these rates are tied to the strong enhancement of
transport observed in chaotic mixing.\cite{Aref1984} Thus, the spatial and
temporal behavior of the finite-time Lyapunov exponents is worthy of study in
its own right, in addition to their time-asymptotic value.

In the present paper we expand upon the work of Tang and
Boozer\cite{Tang1996,Tang1999a,Tang1999b} on the geometrical properties of
finite-time Lyapunov exponents.  We consider an arbitrary smooth dynamical
system, which we take to be a set of ordinary differential equations (ODEs),
but could also be a discrete map.  For such a system, the solution of the
equations of motion is a smooth coordinate transformation (diffeomorphism)
from the space of initial conditions (the Lagrangian coordinates, which label
a trajectory) to the state of the system at a later time (the Eulerian
coordinates).  The coordinate transformation is smooth even when the system is
chaotic, though it then becomes extremely complicated.

To define Lyapunov exponents, it is necessary to define a norm on the space of
our dynamical system, that is, a prescription for measuring distances.  This
norm is used to quantify the exponential separation of trajectories.  The
usual choice is the Euclidean norm, a mathematical embodiment of what we
intuitively think of as distance.  The Euclidean norm is also the simplest
realization of a flat space.  In essence, flatness is an intrinsic property of
a space that reflects the commutativity of partial derivatives, a fact that is
not so trivial in general coordinate systems.  Flatness is taken for granted
in most numerical and laboratory experiments, as reflected by the use of the
Euclidean norm.  When a space is not flat, it is said to have curvature; an
example is a two-dimensional sphere embedded in three-dimensional space, used
in some geophysical models where the atmosphere or the ocean is regarded as
very thin.  But presently we deal only with flat systems.

A remarkable result of differential geometry is the existence and uniqueness
of the Riemann curvature tensor, which vanishes if and only if the the
underlying space is flat (See \secref{Riemann}).  The Riemann tensor plays a
central role in general relativity, where the curvature of space determines
the gravitational forces on matter and light rays.  Here, we shall only use
curvature as a geometrical property; we make no use of general relativity.

When the Riemann tensor vanishes in one coordinate system (usually, in
Cartesian coordinates with the Euclidean norm), it also vanishes in any other
coordinate system reached by a smooth transformation.\cite{truetensor}
In particular, in a smooth dynamical
system the transformation from Lagrangian to Eulerian coordinates leaves a
vanishing Riemann tensor invariant.  When the system is chaotic, so that
trajectories exhibit exponential separation, some terms in the Riemann tensor
exhibit exponential growth, while others decrease exponentially.  Because the
Riemann tensor vanishes identically, these terms have to balance each other;
this requires the finite-time Lyapunov exponents and their associated
characteristic directions to satisfy certain differential relations, or
constraints, as was demonstrated for two-dimensional systems by Tang and
Boozer.\cite{Tang1996} Under more general conditions and using the orthonormal
basis form of the metric tensor,\cite{Wald} we rederive their constraint and
analyze its rate of convergence carefully.  We also extend the calculation to
three dimensions, confirming a constraint conjectured by Tang and
Boozer,\cite{Tang1999b} as well as deriving new ones.  We support our
theoretical claims with numerical simulations on two- (oscillating convection
rolls\cite{Solomon1988}) and three-dimensional flows ($ABC$ flow\cite{STF} and
the Lorenz model\cite{Lorenz1963}). We also comment on the possible physical
applications of the constraints, specifically to the advection-diffusion
equation and the dynamo problem.

The outline of the paper is as follows: in \secref{diffgeom} we review some
basic concepts of differential geometry, as applied to dynamical systems.  In
\secref{Riemann} we focus our attention on the Riemann curvature tensor and
its properties.  Then, in \secref{twoDsystems} we examine the nature of the
vanishing curvature requirement in two dimensions.  We obtain a constraint
relating the derivatives of the smallest Lyapunov exponent and the stable
direction vector.  In \secref{threeDsystems} we carry out the same procedure
in three dimensions, deriving three new constraints.  We find that a
combination of two of these constraints yields the same differential relation
as in the two-dimensional case; the persistence of this constraint in three
dimensions was conjectured by Tang and Boozer,\cite{Tang1999b} who were guided
by numerical investigation of three-dimensional volume-preserving maps.
Finally, in \secref{discussion} we summarize our results, and discuss possible
applications of the constraints.

\section{Differential Geometry and Dynamical Systems}
\seclabel{diffgeom}

In this section we describe the point of view that the solution of a set of
differential equations is a coordinate transformation from the set of initial
conditions to the state of the system at a later time.  In that sense, the
time evolution is a map of the fluid domain onto itself.  Moreover, if we
assume that the dynamics are ``smooth,'' the properties of the coordinate
transformation are described by differential geometry.  An analysis of these
properties is the focus of this paper.

We consider a prescribed smooth two- or three-dimensional vector field given
by~$\vel(\xv,\time)$.  For example,~$\vel(\xv,\time)$ could be a fluid flow
obtained by a solution of the Navier--Stokes equation, but in general the
vector field does not have to correspond to such a fluid flow: it can
represent the right-hand side of any smooth system of ordinary differential
equations.  However, because of the convenient framework provided by a fluid
flow, we often refer to the phase space domain of~$\xv$ as ``the fluid'' and
to infinitesimal elements of phase space as ``fluid elements''.

The \emph{Eulerian} coordinates, $\xv$, denote the position of a point fixed
in the ``laboratory'' frame.  The trajectory of an infinitesimal element of
phase space (also called a fluid element by analogy with the fluid-dynamical
case) in Eulerian coordinates, $\xv$, satisfies
\begin{equation}
	\frac{\pd\xv(\lagrcv,\time)}{\pd\time} =
		\vel(\xv(\lagrcv,\time),\time)\,,
	\qquad \xv(\lagrcv,0) = \lagrcv,
	\eqlabel{lagrcdef}
\end{equation}
where $\lagrcv$ are the \emph{Lagrangian} coordinates that label fluid
elements: they are coordinates that describe a frame moving with the fluid.
We have made the usual choice of taking as initial condition $\xv(\lagrcv,0) =
\lagrcv$, which says that fluid elements are labeled by their initial
position; thus, Eulerian and Lagrangian coordinates coincide at $\time=0$.

The solution $\xv = \xv(\lagrcv,\time)$ to Eq.~\eqref{lagrcdef} is the
transformation from Lagrangian ($\lagrcv$) to Eulerian ($\xv$) coordinates.
For a chaotic flow, this transformation gets extremely complicated as time
evolves.  However, since we have assumed that $\vel$ is a smooth vector field,
the transformation is always differentiable, so that the Jacobian matrix
$\pd\x^i/\pd\lagrc^j$ is well-defined.  This allows us to use the tools of
differential geometry on the transformation $\xv(\lagrcv,\time)$.

Since our focus is on the rate of separation of neighboring trajectories, as
characterized by the Lyapunov exponents, we need to define a way of measuring
distances in the Eulerian coordinates, \ie, a \emph{norm} for vectors in the
space.  The norm we use is the Euclidean norm, defined
by~\hbox{$\|{\mathbf{w}}\|^2 \ldef \sum_{i=1}^{\sdim}w^i\,w^i$}, where~$\sdim$
is the dimension of the space.  With this norm, the distance~$ds$ between two
infinitesimally separated trajectories~$\xv(\lagrcv,\time)$
and~\hbox{$\xv(\lagrcv + d\lagrcv,\time) = \xv(\lagrcv,\time) +
d\xv(\lagrcv,\time)$} is
\begin{equation*}
	ds^2 = \|d\xv\|^2 = \sum_{\ell}^\sdim d\x^\ell\,d\x^\ell.
\end{equation*}
We use the chain rule to write this distance in terms of the initial separation~$d\lagrcv$,
\begin{equation*}
	ds^2 = \sum_{\ell,i,j}^\sdim
		\l(\frac{\pd\x^\ell}{\pd\lagrc^i}\,d\lagrc^i\r)
		\l(\frac{\pd\x^\ell}{\pd\lagrc^j}\,d\lagrc^j\r)
		= \sum_{i,j}^\sdim\metric_{ij}\,d\lagrc^i d\lagrc^j\,,
\end{equation*}
where
\begin{equation}
	\metric_{ij}(\lagrcv,\time) \ldef
		\sum_{\ell=1}^\sdim\frac{\pd\x^\ell}{\pd\lagrc^i}\,
		\frac{\pd\x^\ell}{\pd\lagrc^j}.
	\eqlabel{metricdef}
\end{equation}
In the terminology of differential geometry, Eq.~\eqref{metricdef} is the flat
metric tensor $\flatmetric_{ij}$ of Eulerian space transformed to Lagrangian
coordinates.  The metric $\metric_{ij}$ is a symmetric, positive-definite
matrix that tells us the distance between two infinitesimally separated points
in Lagrangian space.  Because $\metric_{ij}$ is symmetric and
positive-definite, it has $\sdim$ positive real eigenvalues, which we denote
by $\geigen_\mu(\lagrcv,\time)$, with corresponding orthonormal eigenvectors
$\wuv_\mu(\lagrcv,\time)$; we can then write the metric in the diagonal form
\begin{equation}
	\metric_{ij}(\lagrcv,\time)
		= \sum_{\mu=1}^\sdim\geigen_\mu\,(\wu_\mu)_i\,
			(\wu_\mu)_j,
	\eqlabel{diagmetric}
\end{equation}
where~$(\wu_\mu)_i$ is the~$i$th component of~$\wuv_\mu$.  We assume without
loss of generality that~\hbox{$\geigen_\mu \ge \geigen_{\mu+1}$}
for~\hbox{$\mu=1,\ldots,\sdim-1$}.  The eigenvectors~$\wuv_\mu$ define
directions of initial separations for which neighboring fluid elements are
converging or diverging, with~$\geigen_\mu^{1/2}(\lagrcv,\time)$ the distance
separating them.  It is the rate of exponential growth of these distances that
defines the \emph{finite-time Lyapunov exponents} (or characteristic
exponents)
\begin{equation}
	\lyapexp_\mu(\lagrcv,\time) \ldef \frac{1}{2\time}\,
		\ln \geigen_\mu(\lagrcv,\time).
	\eqlabel{lyapexpdef}
\end{equation}
(The factor of $1/2$ enters because $ds^2$ is the square of the distance.)  In
the limit as~$\time\rightarrow\infty$, the finite-time Lyapunov exponents
converge to the true Lyapunov exponents,~$\lyapexpinfty_\mu$, which are
independent of~$\time$ and $\lagrcv$.\cite{indeplyap}
A flow is said to be \emph{chaotic} if at least
one of its Lyapunov exponents converges to a positive value.  However, this
convergence is very slow (logarithmic in time): in practical applications we
are almost always dealing with finite-time Lyapunov exponents.  For the
remainder of this paper, ``Lyapunov exponents'' refers to the finite-time
exponents, unless explicitly denoted as infinite-time.  Whereas the Lyapunov
exponents converge very slowly, in a chaotic region the characteristic
directions~$\wuv_\mu(\lagrcv,\time)$ (the eigenvectors of~$\metric_{ij}$)
converge rapidly (exponentially fast) to their time-asymptotic
value,~$\wuv_\mu^\infty(\lagrcv)$.\cite{Greene1987,Goldhirsch1987}

In the chaotic case, we assume that $\lyapexp_{1} > 0$ ($\geigen_{1} > 1$) and
$\lyapexp_{\sdim} < 0$ ($\geigen_{\sdim} < 1$), so that there is at least one
exponentially expanding and one exponentially contracting direction.  These
are referred to as the unstable and stable directions, respectively.  In three
dimensions, the intermediate exponent~$\lyapexp_{2}$ can have either sign.
For bounded, autonomous flows (where the velocity field $\vel$ does not depend
explicitly on time), the intermediate Lyapunov exponent must converge to zero
in the infinite-time limit.\cite{Eckmann1985} For simplicity, we shall often
use~\hbox{$\mu=\{\edir,\mdir,\sdir\}$} to mean~\hbox{$\mu=\{1,2,3\}$}, and
write~\hbox{$\wuv_\mu = \{\ediruv,\mdiruv,\sdiruv\}$}.  The
letters~$\{\edir,\mdir,\sdir\}$ are abbreviations for ``unstable'',
``middle'', and ``stable'' directions, respectively.  In two dimensions, we
drop the extraneous middle direction and use $\mu=\{\edir,\sdir\}$ to
mean~$\mu=\{1,2\}$, and write~\hbox{$\wuv_\mu = \{\ediruv,\sdiruv\}$}.

The existence of a contracting direction in a chaotic flow follows from the
assumption that the determinant of $\metric_{ij}$
\begin{equation}
	\detmetric \ldef \det\metric_{ij} = \prod_{\mu=1}^\sdim\,
		\geigen_\mu
	\eqlabel{detmetricdef}
\end{equation}
stays bounded with time; otherwise, the volume of fluid elements grows
indefinitely, usually an undesirable situation from the physical standpoint.
However, a bounded~$\detmetric$ is only a sufficient condition for the
existence of a contracting direction, which often exists without this
requirement.  Note that~$\detmetric=1$ for incompressible flows
($\div\vel=0$).

In theory, the numerical procedure for finding the characteristic directions
and exponents is straightforward: we integrate the ordinary differential
equations~\eqref{lagrcdef} with initial conditions~$\lagrcv$, together with
the set of~$\sdim^2$ equations on the tangent space,
\begin{equation}
	\frac{d}{d\time} \l(\frac{\pd\x^i}{\pd\lagrc^j}\r)
	= \sum_{\ell=1}^\sdim
		\frac{\pd\velc^i}{\pd\x^\ell}\,\frac{\pd\x^\ell}{\pd\lagrc^j}.
	\eqlabel{tangentode}
\end{equation}
with initial conditions~\hbox{$\pd\x^i/\pd\lagrc^j = {\delta^i}_j$}.  The
metric is then formed from~$\pd\x^i/\pd\lagrc^j$ using the
definition~\eqref{metricdef}, and its~$\sdim$ eigenvalues~$\geigen_\mu$ and
eigenvectors~$\wuv_\mu$ are obtained via standard numerical techniques.  In
practice, for a chaotic flow, the elements of the Jacobian
matrix~${\pd\x^i}/{\pd\lagrc^j}$ grow exponentially, and one quickly runs into
numerical accuracy problems related to subtracting very large numbers.  Many
methods have been devised to circumvent this
difficulty,\cite{Bennetin1976,Shimada1979,Eckmann1985,Janaki1999} and we shall
use the method of Greene and
Kim\cite{Greene1987,Goldhirsch1987,Christiansen1997}.  Most methods yield a
quantity that differs by a factor proportional to~$1/\time$ from the true
finite-time Lyapunov exponent, and so need to be
corrected.\cite{Goldhirsch1987,Thiffeault2000preprint}

To illustrate the concept of the characteristic directions~$\wuv_\mu$, let
us take the two-dimensional, time-dependent periodic flow,
\begin{equation}
	\velocr
	\ldef \l(-\frac{\pd\psi}{\pd\x^2},\frac{\pd\psi}{\pd\x^1}\r),\qquad
	\psi(\xv,\time) \ldef A\,k^{-1} (\sin k \x^1
		+ \epsilon \cos\omega \time\,\cos k \x^1) \sin\pi \x^2.
	\eqlabel{ocr}
\end{equation}
This flow is incompressible ($\div\vel=0$, so that~$\detmetric=1$).  It is
used by many authors (see for example Refs.~\onlinecite{Solomon1988}
and~\onlinecite{Wiggins}) to model a periodic set of oscillating convection
rolls.  The parameters are the amplitude of the rolls, $A$, the relative
amplitude of the oscillations, $\epsilon$, the aspect ratio, $k$, and the
frequency of oscillations, $\omega$.  When $\omega=0$, the flow is steady, and
the Lagrangian trajectories are nonchaotic.  This is true in general of any
two-dimensional steady flow.\cite{Eckmann1985}

Figure~\ref{fig:oscrollsman} shows the field of stable directions
$\sdiruv^\infty$ for the oscillating rolls, Eq.~\eqref{ocr}.  The two lines in
the figure are portions of stable manifolds, obtained by integrating the set
of~$\sdim$ ordinary differential equations
\begin{equation*}
	\frac{d\lagrcv}{d\tau} = \sdiruv^\infty(\lagrcv),
\end{equation*}
where $\tau$ is the arc length along the manifold.  In a chaotic flow, the
stable manifold comes close to every point in an ergodic
region.\cite{Eckmann1985} In \figref{ABC522sman} we show a portion of stable
manifold for the well-known $ABC$ flow [Eq.~\eqref{ABCflow}], which we discuss
further in \secref{threeDsystems}.  The interpretation of the stable manifold
is as follows: if two infinitesimally close initial conditions lie on that
manifold, their trajectories converge exponentially at a
rate~$\lyapexp_\sdir$.

Analogous to the stable manifold, there is an unstable manifold corresponding
to $\ediruv^\infty$ (and, in three dimensions, a manifold corresponding
to~\hbox{$\mdiruv^\infty = \sdiruv^\infty\times\ediruv^\infty$}).  The
unstable manifold and its relation to mixing have been extensively described
in Refs.~\onlinecite{Giona1999} and~\onlinecite{Cerbelli2000}; an important
result is that material lines (\ie, a streak of dye) in a closed flow tends to
trace out the unstable manifold.

\section{The Riemann Curvature Tensor}
\seclabel{Riemann}

We now apply some well-known results of differential geometry to the metric
obtained in~\secref{diffgeom}.  The aim is to write the Riemann curvature
tensor, which characterizes the flatness of the space, in the basis where the
metric tensor in Lagrangian coordinates is diagonal, as defined by
Eq.~\eqref{diagmetric}.  In \secreftwo{twoDsystems}{threeDsystems}, we will
use the invariance property of the Riemann tensor and the chaotic nature of
the flow to establish constraints on Lyapunov exponents and characteristic
directions.  We only present the properties of the Riemann tensor that are
essential to the argument.  For more complete discussions and proofs, we refer
the reader to standard textbooks, such as
Refs.~\onlinecite{Schutz,Weinberg,Wald}.

Differential geometry tells us that if a metric describes a flat space, then
its Riemann curvature tensor
\begin{equation}
	{\Rcurv^m}_{ijk} \ldef \Christoffel^m_{ji,k} - \Christoffel^m_{ki,j}
		+ \sum_{\ell=1}^\sdim
			\Christoffel^m_{k\ell}\,\Christoffel^\ell_{ji}
		- \sum_{\ell=1}^\sdim
			\Christoffel^m_{j\ell}\,\Christoffel^\ell_{ki},
	\eqlabel{Riemanncurv}
\end{equation}
must vanish in every coordinate system.  (A subscript~$k$ following a comma
denotes a derivative with respect to the~$k$th coordinate.)  The Christoffel
symbols $\Christoffel$ involve derivatives of the metric,
\begin{equation}
	\Christoffel^i_{jk} \ldef \half
	\sum_{\ell=1}^\sdim\metric^{i\ell}\l( \metric_{\ell j,k} +
	\metric_{\ell k,j} - \metric_{jk,\ell}\r),
	\eqlabel{Christoffel}
\end{equation}
where~$\metric^{i\ell}$ is the matrix inverse of~$\metric_{i\ell}$.  The
vanishing of the Riemann tensor is equivalent to requiring that second
covariant derivatives commute in every coordinate system, a result that is
taken for granted in flat space because there covariant derivatives are the
same as ordinary derivatives.  Covariant derivatives can be thought of as
derivatives that take into account a possible spatial dependence of the local
coordinate frame.

The Riemann tensor satisfies a number of symmetry properties, so that its
total number of independent components is~$\sdim^2(\sdim^2-1)/12$.  In three
dimensions, it has six independent components, equivalent to the independent
components of the Ricci tensor~\hbox{$\Ricci_{ik} \ldef
\sum_{j=1}^\sdim{\Ricci^j}_{ijk}$}, which is symmetric.  In two dimensions,
the Riemann tensor has one independent component, equivalent the Ricci
scalar~\hbox{$\Ricci \ldef \sum_{i,k=1}^\sdim\metric^{ik}\,{\Ricci}_{ik}$}. In
this paper we restrict ourselves to the two- and three-dimensional cases, so
we only make use of the Ricci scalar and tensor.

We are interested in finding the components of the Ricci tensor in a frame
aligned with the local characteristic directions (the eigenvectors
of~$\metric_{ij}$, which are orthonormal).  The calculation of the Riemann
tensor in such an orthonormal basis\cite{orthobasis}
simplifies considerably, as described by
Wald.\cite{Wald}  We cite the important results here.

Consider a set of~$\sdim$ orthonormal vectors~$\w_\mu$,
\begin{equation}
	\sum_{i=1}^\sdim
		\,{(\w_{\mu})}^i\,{(\w_{\nu})}_i = \delta_{\mu\nu},\qquad
	\sum_{\mu=1}^\sdim\,{(\w_{\mu})}^i\,{(\w_{\mu})}_j = {\delta^i}_j,
	\eqlabel{orthobasis}
\end{equation}
where~${(\w_{\mu})}^i$ and~${(\w_{\mu})}_i$ are related by the metric
via~${(\w_{\mu})}_i=\sum_j\metric_{ij}\,{(\w_{\mu})}^j$.  In such a basis, the
Ricci tensor is
\begin{equation}
	\Ricci_{\mu\nu} = \sum_{i,\sigma}\covder_i\bigl[
		{(\w_\sigma)}^i\,\Riccirotc_{\mu\nu\sigma}
		- {(\w_\mu)}^i\,\Riccirotc_{\sigma\nu\sigma}\bigr]
	+ \sum_{\sigma,\tau}\l(
	\Riccirotc_{\sigma\nu\sigma}\,\Riccirotc_{\tau\tau\mu}
	- \Riccirotc_{\sigma\tau\mu}\,\Riccirotc_{\tau\nu\sigma}
	\r),
	\eqlabel{Ricciortho}
\end{equation}
where the \emph{Ricci rotation coefficients} are
\begin{equation}
	\Riccirotc_{\sigma\mu\nu} \ldef \sum_{i,j}\,
		(\w_\sigma)^i\,(\w_\mu)^j\,\covder_i\,(\w_\nu)_j.
	\eqlabel{Riccirotcdef}
\end{equation}
We have that~$\Riccirotc_{\sigma\mu\nu} = -\Riccirotc_{\sigma\nu\mu}$, so
there are~$\sdim^2(\sdim-1)/2$ independent~$\Riccirotc$'s (two in two
dimensions, nine in three).  The symbol~$\covder_i$ denotes the covariant
derivative with respect to the~$i$th coordinate.  For our
purposes,~$\covder_i$ can be taken to be an ordinary
derivative,\cite{Riccirotc}
except when it appears as a divergence, where an extra factor involving the
determinant of the metric is required:
\begin{equation}
	\sum_i\,\covder_i\,V^i = \sum_i\,\frac{1}{\sqrt{\detmetric}}\,
		\frac{\pd}{\pd\lagrc^i}\,(\sqrt{\detmetric}\,\,\V^i),
	\eqlabel{covdiv}
\end{equation}
so for incompressible flows there is no practical difference.

The Ricci rotation coefficients carry information similar to the Christoffel
symsbols~$\Christoffel$, but are defined in terms of the orthonormal basis
vectors~$\wv_\mu$ instead of derivatives of the metric.

Summing over the indices of Eq.~\eqref{Ricciortho}, we obtain the Ricci
scalar,
\begin{equation}
	\Ricci = \sum_{\mu}\Ricci_{\mu\mu} = 2\sum_{i,j,\sigma}
		\covder_i\bigl[
		{(\w_\sigma)}^i\,\covder_j{(\w_\sigma)}^j\bigr]
	- \sum_{\sigma,\tau,\mu}\l(
	\Riccirotc_{\sigma\sigma\mu}\,\Riccirotc_{\tau\tau\mu}
	- \Riccirotc_{\sigma\tau\mu}\,\Riccirotc_{\tau\sigma\mu}
	\r),
	\eqlabel{Ricciscal}
\end{equation}
where we have made use of the identity
\begin{equation*}
	\sum_\mu\,\Riccirotc_{\mu\mu\sigma} = \sum_i\covder_i(\w_\sigma)^i,
\end{equation*}
easily verified directly from the definition of~$\Riccirotc$,
Eq.~\eqref{Riccirotcdef}.  The Ricci tensor and scalar in an orthonormal
frame, respectively Eqs.~\eqref{Ricciortho} and~\eqref{Ricciscal}, are the
quantities we use in \secreftwo{twoDsystems}{threeDsystems} to derive
constraints on dynamical systems.

Equation~\eqref{orthobasis} states that the vectors~$\wv_\mu$ are orthonormal
with respect to a metric.  We want to use the metric given by
Eq.~\eqref{diagmetric}, so we take
\begin{equation}
	(\w_\mu)_i = \geigen_\mu^{1/2}\,(\wu_\mu)_i, \qquad
	(\w_\mu)^i = \geigen_\mu^{-1/2}\,(\wu_\mu)^i,
	\eqlabel{wdef}
\end{equation}
where, because the underlying space is Euclidean, we have~\hbox{$(\wu_\mu)_i =
(\wu_\mu)^i$}.  It is then clear that~\eqref{orthobasis} is satisfied.  The
definitions~\eqref{wdef} are called the covariant (subscripted) and
contravariant (superscripted) representations of the vectors~$\wv_\mu$.  For a
chaotic system, this definition of~$\wv_\mu$ implies that for long times the
components~$(\w_\sdir)^i \rdef \sdir^i$ and~$(\w_\edir)_i \rdef \edir_i$ are
growing exponentially, and~$(\w_\sdir)_i \rdef \sdir_i$ and~$(\w_\edir)^i
\rdef \edir^i$ are decaying exponentially.  It is these radically different
time-asymptotic behaviors that allows the derivation of the constraints in
\secreftwo{twoDsystems}{threeDsystems}.

\section{Constraints on Two-dimensional Systems}
\seclabel{twoDsystems}

In \secref{Riemann}, we derived a form for the Ricci tensor expressed in a
basis aligned with the characteristic directions~$\wuv_\mu$ of the flow.  As
mentioned in~\secref{Riemann}, the symmetries of the Riemann curvature tensor
imply that in two dimensions it has only one independent component, given by
the Ricci scalar, Eq.~\eqref{Ricciscal}.  In two dimensions, the terms in the
Ricci scalar quadratic in~$\Riccirotc$ cancel:
\begin{equation*}
\sum_{\sigma,\tau,\mu}\l(
	\Riccirotc_{\sigma\sigma\mu}\,\Riccirotc_{\tau\tau\mu}
	- \Riccirotc_{\sigma\tau\mu}\,\Riccirotc_{\tau\sigma\mu}\r)
	= 
	\Riccirotc_{221}\,\Riccirotc_{221}
	+ \Riccirotc_{112}\,\Riccirotc_{112}
	- \Riccirotc_{221}\,\Riccirotc_{221}
	- \Riccirotc_{112}\,\Riccirotc_{112} = 0.
\end{equation*}
Hence, the Ricci scalar reduces to the simple form
\begin{equation}
	\Ricci = 2\sum_{i,j,\sigma=1}^2
		\covder_i\,\l[(\w_{\sigma})^i\,
		\covder_j(\w_{\sigma})^j\r].
	\eqlabel{Ricciscalar2d}
\end{equation}
As mentioned in \secref{diffgeom}, we are assuming the underlying space is
flat, so that~\hbox{$\Ricci\equiv 0$} always.  Equation~\eqref{Ricciscalar2d}
is essentially the same expression as derived in Ref.~\onlinecite{Tang1996},
but rewritten in a more transparent form and allowing for compressibility of
the vector field~$\vel(\xv,\time)$ (because here the derivatives~$\covder_i$
are covariant).  Notice that the Lyapunov exponents enter the Ricci scalar
through the definition of~$(\w_\sigma)^i$, Eq.~\eqref{wdef}, as
$\geigen_\sigma^{-1/2} = \exp(-\lyapexp_\sigma\,\time)$.

As a direct demonstration of the vanishing of the Ricci scalar for the
nonchaotic case, let us take the flow $\vel(\x_1,\x_2) = (f(\x_2),0)$, a shear
flow of velocity~$f(\x_2)$ along the~$\x_1$ direction.  For this special case,
Eq.~\eqref{lagrcdef} can be explicitly solved, and the Lagrangian trajectories
are given by
\begin{equation*}
	\x_1 = \lagrc_1 + \time\,f(\lagrc_2),\qquad
	\x_2 = \lagrc_2.
\end{equation*}
The corresponding metric tensor is
\begin{equation*}
\metric_{ij} = 	\sum_{\ell}\frac{\pd\x^\ell}{\pd\lagrc^i}\,
	\frac{\pd\x^\ell}{\pd\lagrc^j}
= \begin{pmatrix}
	1 & \time\,f'(\lagrc_2) \\
	\time\,f'(\lagrc_2) & 1 + \time^2\,f'(\lagrc_2)^2
\end{pmatrix}.
\end{equation*}
The eigenvalues and eigenvectors of $\metric$ are then straightforward to
calculate.  Direct insertion into formula~\eqref{Ricciscalar2d} for the Ricci
scalar in two-dimensions confirms, after a tedious calculation, that it does
indeed vanish identically.  When the flow is not chaotic, the usefulness of
Eq.~\eqref{Ricciscalar2d} is thus somewhat limited: it is simply an identity
that is always satisfied.  At best, verification that the Ricci scalar
vanishes is a somewhat complicated consistency check.

The real usefulness of Eq.~\eqref{Ricciscalar2d} becomes more apparent when
the Lagrangian trajectories are taken to be chaotic (which in two dimensions
requires a time-dependent flow). It is still true that the Ricci
scalar~\eqref{Ricciscalar2d} vanishes identically, because the coordinate
transformation~$\xv(\lagrcv,\time)$ is differentiable.  However, there is
information to be gained by considering what we call the \emph{curvature
balance}, that is, the details of how the vanishing of the curvature tensor is
realized in the chaotic flow.  We now proceed to examine this balance.

Let us write
\begin{equation}
	\lyapexp_\mu(\lagrcv,\time)
	= \lyapexp_\mu^\infty
		+ \frac{\lyapexpt_\mu(\lagrcv,\time)}{\time},
	\eqlabel{lyapevol}
\end{equation}
where the dimensionless function~$\lyapexpt_\mu(\lagrcv,\time)$
satisfies~\hbox{$\lim_{\time\rightarrow\infty}
\lyapexpt_\mu(\lagrcv,\time)/\time = 0$}.  The choice of the factor
of~$1/\time$ is motivated by the form for the time-evolution of the Lyapunov
exponents derived by Goldhirsch \etal~\cite{Goldhirsch1987} based on a direct
analysis of the differential equations they satisfy.
Essentially,~$\lyapexpt_\mu(\lagrcv,\time)$ is a ``noise'' term of relatively
small amplitude: the dominant time-asymptotic behavior of the finite-time
Lyapunov exponents is as~$1/\time$.  Note that with this
definition~\hbox{$\covder_i\lyapexp_\mu\,\time = \covder_i\lyapexpt_\mu$}.

Having posed the form~\eqref{lyapevol} for the finite-time Lyapunov exponents,
we can rewrite~\eqref{Ricciscalar2d} as
\begin{equation}
\Ricci =
2\sum_{i,j,\sigma=1}^2
\ee^{-2\lyapexpinfty_\sigma\,\time}\,
	\covder_i\Bigl[\ee^{-\lyapexpt_\sigma}\,(\wu_\sigma)^i\,
	\covder_j\bigl(\ee^{-\lyapexpt_\sigma}\,(\wu_\sigma)^j\bigr)\Bigr].
	\eqlabel{Ricciscalartimefac}
\end{equation}
Now we can clearly see the asymptotic time dependence of each term for
large~$\time$:
\begin{equation}
\Ricci =
2\ee^{-2|\lyapexpinfty_{\edir}|\,\time}\,
		\covder_i\Bigl[\ee^{-\lyapexpt_{\edir}}\,\ediru^i\,
		\covder_j\bigl(
			\ee^{-\lyapexpt_{\edir}}\,\ediru^j\bigr)\Bigr]
+ 2\ee^{+2|\lyapexpinfty_{\sdir}|\,\time}\,
		\covder_i\Bigl[\ee^{-\lyapexpt_{\sdir}}\,\sdiru^i\,
		\covder_j\bigl(
			\ee^{-\lyapexpt_{\sdir}}\,\sdiru^j\bigr)\Bigr]
	\eqlabel{Riccitwodexpand}
\end{equation}
We have put absolute values around the exponents to exhibit the sign of the
exponential, following our assumption that there is an expanding (positive)
exponent,~$\lyapexpinfty_{\edir}$, and a contracting (negative)
exponent,~$\lyapexpinfty_{\sdir}$.

Since~$\Ricci$ vanishes identically, either the two terms
in Eq.~\eqref{Riccitwodexpand} both vanish identically (as is the case for
purely hyperbolic systems such as Arnold's cat map\cite{Tang1996}), or they
must balance each other, growing at the same exponential rate~$2\Riccigr$:
\begin{equation}
\Ricci =
2\ee^{2\Riccigr\,\time}
	\l\{\ee^{-2(|\lyapexpinfty_{\edir}| + \Riccigr)\,\time}\,
		\covder_i\Bigl[\ee^{-\lyapexpt_{\edir}}\,\ediru^i\,
		\covder_j\bigl(
			\ee^{-\lyapexpt_{\edir}}\,\ediru^j\bigr)\Bigr]
	+ \ee^{+2(|\lyapexpinfty_{\sdir}| - \Riccigr)\,\time}\,
		\covder_i\Bigl[\ee^{-\lyapexpt_{\sdir}}\,\sdiru^i\,
		\covder_j\bigl(
			\ee^{-\lyapexpt_{\sdir}}\,\sdiru^j\bigr)\Bigr]\r\}.
	\eqlabel{Ricciscalgr}
\end{equation}
In other words, the growth rate~$\Riccigr$ is defined by the requirement that
each term inside the braces in Eq.~\eqref{Ricciscalgr} be of order one
asymptotically.  Thus, for large time we must have
\begin{align}
	\covder_i\Bigl[\ee^{-\lyapexpt_{\edir}}\,\ediru^i\,
		\covder_j\bigl(\ee^{-\lyapexpt_{\edir}}\,\ediru^j
		\bigr)\Bigr]
	&\sim \exp(+2(|\lyapexpinfty_{\edir}| + \Riccigr)\,\time),
	\eqlabel{edir2dgrowthasymp}\\
	\covder_i\Bigl[\ee^{-\lyapexpt_{\sdir}}\,\sdiru^i\,
		\covder_j\bigl(\ee^{-\lyapexpt_{\sdir}}\,\sdiru^j
		\bigr)\Bigr]
	&\sim \exp(-2(|\lyapexpinfty_{\sdir}| - \Riccigr)\,\time).
	\eqlabel{sdir2dgrowthasymp}
\end{align}
These growth rates apply even if~$\sqrt\detmetric$ grows or decays
exponentially in time, since because of Eq.~\eqref{covdiv} it appears both in
the numerator and the denominator.  If~\hbox{$\Riccigr >
|\lyapexpinfty_{\sdir}|$}, then both~\eqref{edir2dgrowthasymp}
and~\eqref{sdir2dgrowthasymp} \emph{grow} exponentially.  If~\hbox{$\Riccigr <
|\lyapexpinfty_{\sdir}|$}, for~\hbox{$\time\gg (|\lyapexpinfty_{\sdir}| -
\Riccigr)^{-1}$} we have
\begin{equation}
	\covder_i\Bigl[\ee^{-\lyapexpt_{\sdir}}\,\sdiru^i\,
		\covder_j\bigl(
		\ee^{-\lyapexpt_{\sdir}}\,\sdiru^j\bigr)\Bigr] = 0
	\eqlabel{sdir2dconstraint}
\end{equation}
to exponential accuracy in~$\time$.  In practice, we have found only cases
with~\hbox{$-|\lyapexpinfty_{\edir}| \le \Riccigr < |\lyapexpinfty_{\sdir}|$}.
Equation~\eqref{sdir2dconstraint} is thus a differential constraint equation
relating~$\sdiruv$ and~$\lyapexpt_\sdir$.

Note that we do not have to wait for the Lyapunov exponent itself to have
converged for the constraint~\eqref{sdir2dconstraint} to be satisfied: it is
sufficient that the quantity~$|\lyapexp_\sdir| - \Riccigr$ have a definite
sign, which occurs rapidly in practice.

Having derived Eq.~\eqref{sdir2dconstraint}, we now show that it can be
reduced to a simpler constraint.  Let
\begin{equation}
	\Kdivs \ldef \covder_i\bigl(\ee^{-\lyapexpt_{\sdir}}\,\sdiru^i\bigr).
	\eqlabel{Kdivsdef}
\end{equation}
The constraint Eq.~\eqref{sdir2dconstraint} can be written
\begin{equation}
	\frac{d\Kdivs}{d\bar\tau} = -\Kdivs^2,\qquad
	\frac{d}{d\bar\tau} \ldef
		\ee^{-\lyapexpt_{\sdir}}\,\sdiru^i\,\covder_i,
	\eqlabel{Kode}
\end{equation}
where~$\bar\tau$ is a parameter that measures the distance along
an~$\sdiruv$-line, weighed by a density of~$\exp(-\lyapexpt_{\sdir})$ [this
relation is well-defined because~\hbox{$\exp(-\lyapexpt_{\sdir}) > 0$}].  The
ordinary differential equation~\eqref{Kode} has solution
\begin{equation}
	\Kdivs = \frac{\Kdivs_0}{1 + (\bar\tau - \bar\tau_0)\Kdivs_0}\,,
	\eqlabel{Kbadsoln}
\end{equation}
which has a singularity at~$\bar\tau-\bar\tau_0=-\Kdivs_0^{-1}$.  Since all
the quantities that make up~$\Kdivs$ are well-defined for a given point on
the~$\sdiruv$-line, we have to reject the nonvanishing solution to
Eq.~\eqref{Kode}, and instead take~$\Kdivs\equiv0$.  Hence, the constraint
\begin{equation}
	\covder_i\bigl(\ee^{-\lyapexpt_\sdir}\,\sdiru^i\bigr) = 0
	\eqlabel{sdir2dconstraintsimp}
\end{equation}
is necessary and sufficient for Eq.~\eqref{sdir2dconstraint} to be satisfied.
Equation~\eqref{sdir2dconstraintsimp} can also be written
\begin{equation}
	\covder_i\sdiru^i - \sdiru^i\,\covder_i\lyapexpt_\sdir = 0.
	\eqlabel{sdir2dconstraintsimp2}
\end{equation}
This is a compressible version of the constraint that was obtained in
Ref.~\onlinecite{Tang1996}.  If we use~\hbox{$\ln\sqrt{\detmetric} =
\lyapexp_{\edir}\time + \lyapexp_{\sdir}\time$}, we can
rewrite~\eqref{sdir2dconstraintsimp2} as
\begin{equation}
	\divlc\sdiruv
	+ \sdiruv\cdot\gradlc\lyapexp_\edir\,\time = 0,
	\eqlabel{sdirle2dconstraint}
\end{equation}
where the~$\gradlc$ are ordinary derivatives (\ie, non-covariant) and any
direct reference to compressibility effects drops out, but the constraint now
takes the form of a relation between the stable direction, $\sdiruv$, and the
unstable stretching rate, $\lyapexp_{\edir}$.

To illustrate the result, we use the oscillating convection rolls system,
Eq.~\eqref{ocr}.  We consider the~$\sdiruv$-line starting at~$A$ and ending
at~$B$ in \figref{oscrollsman}, with the same parameter values.  In
\figref{twodslineconverg} we plot the two terms of
Eq.~\eqref{sdirle2dconstraint} along that line for different times; it is
clear that their sum converges rapidly to zero.
Figure~\ref{fig:twodslineconverg} highlights another property of the
constraint~\eqref{sdir2dconstraintsimp2}: if
$\lim_{\time\rightarrow\infty}\covder_i\,\sdiru^i$ exists (which it does in
every case under consideration here), so
must~$\lim_{\time\rightarrow\infty}\sdiru^i\,\covder_i\lyapexpt_{\sdir}$.
\comment{Grrr... would be nice to \emph{prove} that the limit exists.}

To exhibit the rate of exponential convergence to zero of
Eq.~\eqref{sdirle2dconstraint}, we consider the flow
\begin{equation}
	\dot \x^1 = \x^2 + \sin \x^1 \,\sin \x^2,\qquad
	\dot \x^2 = \mu\,\x^1 + \nu\,\x^2 + \cos \x^1 \,\cos \x^2.
	\eqlabel{nonchaotic2d}
\end{equation}
This is a two-dimensional autonomous system, so it cannot formally exhibit
chaos.  Nevertheless, for adequate choice of~$\mu$ and~$\nu$ trajectories
of~\eqref{nonchaotic2d} grow exponentially away from the origin, so that
Lyapunov exponents are well-defined even though they are not tied to any chaos
in the system.  One could also achieve such nonzero exponents with a linear
system, but then the~$\ediruv$- and~$\sdiruv$-lines would be straight, with
vanishing derivatives, so that Eq.~\eqref{sdirle2dconstraint} is satisfied
trivially.  The nonlinear terms in in~\eqref{nonchaotic2d} do not contribute
to~$\div\vel=\nu$.  Figure~\ref{fig:nonchaotic2dconverg} shows the evolution
of~\hbox{$\divlc\sdiruv + \sdiruv\cdot\gradlc\lyapexp_\edir\,\time$} for
parameter values~$\mu=1$,~$\nu=-1$.  The convergence rate is found to
be~\hbox{$-(\lyapexp_\edir + |\lyapexp_\sdir|)$}, corresponding to~$\Riccigr =
-\lyapexp_\edir$ in Eq.~\eqref{sdir2dgrowthasymp}.  This implies from
Eq.~\eqref{edir2dgrowthasymp}
that~$\covder_i(\exp(-\lyapexpt_\edir)\,\ediru^i)$ remains of order one, as
confirmed by numerical calculation (\figref{nonchaotic2dconverg}, bottom).  In
truly chaotic systems, the convergence rate is not so readily obtained because
the coefficient of the exponential has large fluctuations.

The rate of convergence of the nonchaotic system seems to be an extreme case.
The derivatives of the Lyapunov exponents along the unstable direction are
well-behaved and do not grow exponentially, so from
Eq.~\eqref{edir2dgrowthasymp} we must have~\hbox{$\Riccigr =
-\lyapexp_\edir$}.  This leads to an extremely fast convergence rate for the
constraint~\eqref{sdirle2dconstraint}.  For chaotic systems, we find the
derivatives of the Lyapunov exponents along the unstable direction grow
exponentially, so the convergence rate of the constraint is slower (but still
quite fast).

The numerical evaluation of the Lagrangian derivatives
of~$\lyapexp_\mu(\lagrcv,\time)$ and~$\wv_\mu(\lagrcv,\time)$ is not
straightforward.  Finite-differencing (by starting two trajectories very close
to each other) does not work well because of potentially very steep variations
along the unstable direction.  Also, it is difficult to guarantee that the two
trajectories are in the same chaotic region when they are very close to region
boundaries.  A better method is to take the Lagrangian derivative of the set
of equations used to find the Lyapunov exponents and eigenvectors themselves,
which yields a set of ordinary differential equations for those derivatives.
This is related to the method used by Tang and Boozer,\cite{Tang1996} who used
the Lagrangian derivative of the linearized system~\eqref{tangentode}
directly.  The method is unstable, but since the quantities under
consideration converge quickly (the Lyapunov exponents are not sought, only
their derivatives), reasonable accuracy can be achieved.  In this paper, we
have used a differentiated version of the continuous Gram--Schmidt
orthonormalization method of Goldhirsch~\etal\cite{Goldhirsch1987} with a
stabilizing factor.\cite{Christiansen1997,Thiffeault2000preprint}  The method
of Goldhirsch~\etal\ was derived independently by Greene and
Kim,\cite{Greene1987} including the stabilizing factor.

\section{Constraints on Three-dimensional Systems}
\seclabel{threeDsystems}

In three spatial dimensions, the six independent components of the Riemann
curvature tensor are embodied in the Ricci tensor (see \secref{Riemann}).  As
opposed to the two-dimensional case, the terms quadratic in the~$\Riccirotc$'s
in the Ricci tensor, Eq.~\eqref{Ricciortho}, do not vanish.

The three diagonal elements of the Ricci tensor can be written as
\begin{multline}
\Ricci_{\tau\tau} =
	\covder_i\l[(\w_\tau)^i(\helic_{\nu\mu}
			- \helic_{\mu\nu})
		+ (\w_\nu)^i\,\helic_{\mu\tau}
		- (\w_\mu)^i\,\helic_{\nu\tau}\r]\\
	+ 2\helic_{\mu\nu}\,\helic_{\nu\mu}
	+ \half\l[
	\l(\helic_{\mu\mu} - \helic_{\nu\nu}\r)^2
		- \helic_{\tau\tau}^2\r]
	\eqlabel{Ricci3Ddiag}
\end{multline}
and the three off-diagonal elements as
\begin{multline}
\Ricci_{\mu\nu} = \covder_i\l[
	(\w_\nu)^i\,\helic_{\nu\tau}
			- (\w_\mu)^i\,\helic_{\mu\tau}
		+ (\w_\tau)^i\,(\helic_{\mu\mu} - \helic_{\nu\nu})
	\r]\\
+ \helic_{\tau\mu}\,\helic_{\nu\tau}
		+ \helic_{\tau\nu}\,\helic_{\mu\tau}
	+ (\helic_{\mu\nu} + \helic_{\nu\mu})
		(\helic_{\mu\mu}
		+ \helic_{\nu\nu}
		- \helic_{\tau\tau}),
	\eqlabel{Ricci3Doffdiag}
\end{multline}
where~$(\mu,\nu,\tau)$ are ordered cyclically (\ie,
\hbox{$(\mu,\nu,\tau) =
\{(\edir,\mdir,\sdir),(\mdir,\sdir,\edir),(\sdir,\edir,\mdir)\}$}) and the
\emph{characteristic helicities} are defined as
\begin{equation*}
	\helic_{\mu\nu} \ldef \frac{1}{\sqrt{\detmetric}}\,
		(\w_\mu)_i\,\varepsilon^{ijk}\,\covder_j(\w_\nu)_k\,.
\end{equation*}
The main result of this section is as follows: if all the terms in
Eq.~\eqref{Ricci3Ddiag} are bounded from above
by~$\exp(2|\lyapexpinfty_{\sdir}|\time)$, then the constraints
\begin{align}
	\ediruv\cdot\curllc\mdiruv - \sdiruv\cdot\gradlc
		\lyapexp_\mdir\,\time
		&= 0,
	\eqlabel{Hemconstr}\\
	\mdiruv\cdot\curllc\ediruv + \sdiruv\cdot\gradlc
		\lyapexp_\edir\,\time
		&= 0,
	\eqlabel{Hmeconstr}\\
	\ediruv\cdot\curllc\ediruv &= 0,
	\eqlabel{Heeconstr}
\end{align}
hold to exponential accuracy for large times.  Taking the difference
between~\eqref{Hmeconstr} and~\eqref{Hemconstr} and using vector identities
yields
\begin{equation}
	\frac{1}{\sqrt{\detmetric}}\,
		\divlc\l(\sqrt{\detmetric}\,\sdiruv\r)
	- \sdiruv\cdot\gradlc\lyapexp_\sdir\,\time
	\longrightarrow 0,
	\eqlabel{sdir2dconstraintsimp2b}
\end{equation}
the same constraint as for the two-dimensional case,
Eq.~\eqref{sdir2dconstraintsimp2}.  The
constraint~\eqref{sdir2dconstraintsimp2b} was observed numerically for
three-dimensional volume-preserving maps by Tang and Boozer.\cite{Tang1999b}
We thus see that Eq.~\eqref{sdir2dconstraintsimp2b} is a consequence of two
new, separate constraints.  The third constraint, Eq.~\eqref{Heeconstr}, is
different in nature than the previous ones, since it involves no Lyapunov
exponents.

If we ``lift'' a two-dimensional system to three dimensions by simply adding a
third dimension corresponding to~\hbox{$\mdiruv = \sdiruv\times\ediruv =
\text{constant}$}, the constraint~\eqref{Hemconstr} is satisfied trivially
because~\hbox{$\lyapexp_{\mdir} = 0$} and~$\mdiruv$ is independent of
position.  The constraint~\eqref{Hmeconstr} can be equated
to~\eqref{sdir2dconstraintsimp2}, the two-dimensional constraint,
and~\eqref{Heeconstr} is trivially satisfied because~$\curllc\ediruv$ is in
the~$\mdiruv$ direction.  Thus, as required for consistency, we recover the
two-dimensional constraint from the three-dimensional ones.

We now proceed with the derivation of the
constraints~\eqref{Hemconstr}--\eqref{Hmeconstr}.  The argument is similar to
the two-dimensional case, but is complicated by the possibility that the terms
in Eq.~\eqref{Ricci3Ddiag} and~\eqref{Ricci3Doffdiag} grow at different rates,
since there are more than two terms in the expression.  There are also six
independent components to the Ricci tensor, as opposed to just one in the
two-dimensional case.

Define
\begin{equation}
	\helict_{\mu\nu} \ldef
	(\wu_\mu)_i\,\varepsilon^{ijk}\,\covder_j(\wu_\nu)_k
	- \sum_\tau
	\varepsilon_{\mu\nu\tau}(\wu_\tau)^j\,\covder_j\lyapexpt_\nu
	\eqlabel{helictdef}
\end{equation}
where~$\lyapexpt_\nu$ is defined in~\eqref{lyapevol}, so
that~\hbox{$\helic_{\mu\mu} =
(\geigen_{\mu}/\sqrt{\detmetric})\,\helict_{\mu\mu}$}
and~\hbox{$\helic_{\mu\nu} =
\geigen_{\tau}^{-1/2}\,\helict_{\mu\nu}$},~$\mu\ne\nu$.  With this definition
we have separated the long-time behavior of~$\helic_{\mu\nu}$ that is due to
the infinite-time Lyapunov exponents.

We focus first on the diagonal components, given by Eq.~\eqref{Ricci3Ddiag}.
Let~$2\Riccigr_{\tau\tau}$ be the asymptotic growth rate of the
fastest-growing term (or terms) in~\eqref{Ricci3Ddiag}.  (In two dimensions,
there was only one common growth rate~$\Riccigr$ for the two terms
in~$\Ricci$, since they had to cancel.)  Factoring
out~$\exp(2\Riccigr_{\tau\tau}\time)$ from Eq.~\eqref{Ricci3Ddiag}, we obtain
\begin{multline}
\Ricci_{\tau\tau} = \ee^{2\Riccigr_{\tau\tau}\time}
	\biggl\{\covder_i\Bigl[\ee^{-2(\lyapexp_\tau
		+ \Riccigr_{\tau\tau})\time}
		(\wu_\tau)^i(\helict_{\nu\mu}
			- \helict_{\mu\nu})
		+ \ee^{-2(\lyapexp_\nu + \Riccigr_{\tau\tau})\time}
			(\wu_\nu)^i\,\helict_{\mu\tau}\\
		- \ee^{-2(\lyapexp_\mu + \Riccigr_{\tau\tau})\time}
			(\wu_\mu)^i\,\helict_{\nu\tau}\Bigr]
	+ 2\ee^{-2(\lyapexp_\tau + \Riccigr_{\tau\tau})\time}\,
		\helict_{\mu\nu}\,\helict_{\nu\mu}\\
	+ \half\l[\l(\ee^{-(\lyapexp_\nu + \lyapexp_\tau - \lyapexp_\mu
		+ \Riccigr_{\tau\tau})\time}\,\helict_{\mu\mu}
	- \ee^{-(\lyapexp_\mu + \lyapexp_\tau - \lyapexp_\nu
		+ \Riccigr_{\tau\tau})\time}\,\helict_{\nu\nu}\r)^2
	- \ee^{-2(\lyapexp_\mu + \lyapexp_\nu - \lyapexp_\tau
		+ \Riccigr_{\tau\tau})\time}\,
\helict_{\tau\tau}^2\r]\biggr\},
\end{multline}
where, by the definition of~$\Riccigr_{\tau\tau}$, all the terms inside the
braces are asymptotically either of order one or decreasing exponentially in
time.  We are interested in the terms that have a positive growth rate in the
exponential; for such terms to remain of order one or less the coefficient of
the exponential has to go to zero exponentially, as was the case in two
dimensions.

In each~$\Ricci_{\tau\tau}$ there is a term with
coefficient~$\exp(-2(\lyapexpinfty_{\sdir} + \Riccigr_{\tau\tau})\time)$:
if~\hbox{$\Riccigr_{\tau\tau} < |\lyapexpinfty_{\sdir}|$} then that
coefficient grows exponentially at a
rate~\hbox{$\exp(2(|\lyapexpinfty_{\sdir}| - \Riccigr_{\tau\tau})\time)$}.
This implies that the coefficient of the exponential decreases at least as
fast as~\hbox{$\exp(-2(|\lyapexpinfty_{\sdir}| - \Riccigr_{\tau\tau})\time)$}.
Thus, from each diagonal components~$\Ricci_{\tau\tau}$, we get
\begin{align}
	\covder_i\l[\ee^{-2\lyapexpt_{\sdir}}\sdiru^i\,\helict_{\mdir\edir}\r]
	&< \exp({-2(|\lyapexpinfty_{\sdir}| - \Riccigr_{\edir\edir})\time}),
	\qquad \text{from\ }\Ricci_{\edir\edir}\,;
	\eqlabel{Ree1cst}\\
	\covder_i\l[\ee^{-2\lyapexpt_{\sdir}}\sdiru^i\,\helict_{\edir\mdir}\r]
	&< \exp({-2(|\lyapexpinfty_{\sdir}| - \Riccigr_{\mdir\mdir})\time}),
	\qquad \text{from\ }\Ricci_{\mdir\mdir}\,;
	\eqlabel{Ree2cst}\\
	\covder_i\l[\ee^{-2\lyapexpt_{\sdir}}\sdiru^i\,
		(\helict_{\mdir\edir} - \helict_{\edir\mdir})\r]
		+ 2\helict_{\mdir\edir}\,\helict_{\edir\mdir}
	&< \exp({-2(|\lyapexpinfty_{\sdir}| - \Riccigr_{\sdir\sdir})\time}),
	\qquad \text{from\ }\Ricci_{\sdir\sdir}\,;
	\eqlabel{Ree3cst}
\end{align}
The two constraints~\eqref{Ree1cst} and~\eqref{Ree2cst} can be directly
inserted into~\eqref{Ree3cst}, which then says that the
product~$\helict_{\mdir\edir}\,\helict_{\edir\mdir}$ decreases exponentially
to zero.  Since
\begin{equation*}
	\helict_{\mdir\edir} - \helict_{\edir\mdir} =
	\ee^{\lyapexpt_\sdir}\,\Kdivs,
\end{equation*}
where~$\Kdivs$ is defined in~\eqref{Kdivsdef}, we can use the argument, proved
in the two-dimensional case, that if~\hbox{$\covder_i[\sdir^i\,\Kdivs]=0$},
then~$\Kdivs=0$.  Thus,~\hbox{$\helict_{\mdir\edir} = \helict_{\edir\mdir}$},
and their product goes to zero, so both~$\helict_{\mdir\edir}$
and~$\helict_{\edir\mdir}$ must go to zero.  From the
definition~\eqref{helictdef}, we see that we have just demonstrated the
constraints~\eqref{Hemconstr} and~\eqref{Hmeconstr}, with the assumption
that~\hbox{$\Riccigr_{\tau\tau} < |\lyapexpinfty_\sdir|$}.

Also, in each~$\Ricci_{\tau\tau}$ there is a term with
coefficient~\hbox{$\exp(2(\lyapexpinfty_{\edir} - \lyapexpinfty_{\mdir} -
\lyapexpinfty_{\sdir} - \Riccigr_{\tau\tau})\time)$}.
Since~\hbox{$\lyapexpinfty_\edir > \lyapexpinfty_\mdir$}, the
condition~\hbox{$\Riccigr < |\lyapexpinfty_\sdir|$} is sufficient for the
growth rate to be positive.  This leads directly to
\begin{equation*}
	\helict_{\edir\edir} < \exp(-(\lyapexpinfty_{\edir} -
\lyapexpinfty_{\mdir} + |\lyapexpinfty_{\sdir}| - \Riccigr_{\tau\tau})\time),
\end{equation*}
which proves the last constraint, Eq.~\eqref{Heeconstr}.

There could be other constraints, depending on the relative magnitude of the
Lyapunov exponents and the sign of~$\lyapexpinfty_\mdir$.  However, the
constraints derived here have the broadest applicability and are expected to
be true in general.  The requirement that~\hbox{$\Riccigr <
|\lyapexpinfty_\sdir|$} has been found to hold by a wide margin for every
dynamical system that we have tested.  Given that the Lyapunov exponents enter
the Ricci tensor through~$\geigen_\mu^{-1}$, so that the largest growth rate
is~$2|\lyapexpinfty_\sdir|$, it is reasonable to assume that the terms in the
Ricci scalar could grow no faster than the
timescale~\hbox{$2\Riccigr = 2|\lyapexpinfty_\sdir|$}.

To exhibit the convergence of these quantities, we use the model system
\begin{equation}
	\dot \x^1 = \x^2 + \sin \x^1 \,\sin \x^2 \,\sin \x^3,\qquad
	\dot \x^2 = \x^3 + \cos \x^1 \,\cos \x^2 \,\sin \x^3,\qquad
	\dot \x^3 = \lambda\,\x^1 + \mu\,\x^2 + \nu\,\x^3.
	\eqlabel{nf3}
\end{equation}
These are an extension of the two-dimensional nonchaotic
equations~\eqref{nonchaotic2d} presented earlier.  In the three-dimensional
case, the system is generically chaotic, but can have unbounded trajectories
for certain parameter values.  If we pick an initial condition near the
origin, then initially the nonlinear terms are important, but become less so
as the trajectory gets further away.  Thus, Eqs.~\eqref{nf3} have nontrivial
characteristic manifolds near the origin but have a time-asymptotic form that
is well-behaved (linear exponential growth).
Figures~\ref{fig:LogHumHmuNF3_set02} and~\ref{fig:LogHumHmuNF3_set01} show the
convergence of~\eqref{Hemconstr} and~\eqref{Hmeconstr} for different parameter
values, one set incompressible ($\nu=0$) and the other compressible ($\nu\ne
0$); Figures~\ref{fig:LogHuuNF3_set02} and~\ref{fig:LogHuuNF3_set01} show the
convergence of~\eqref{Heeconstr} for the same parameter values.  For this
simple system, the rate of convergence for all the constraints is found to
be~\hbox{$-2(\lyapexp_\edir - \lyapexp_\mdir)$}.

A more practical example is the $ABC$ flow,
\begin{equation}
	\vel(\xv) = A\,(0,\sin\x^1,\cos\x^1) + B\,(\cos\x^2,0,\sin\x^2)
		+ C\,(\sin\x^3,\cos\x^3,0)
	\eqlabel{ABCflow}
\end{equation}
a sum of three Beltrami waves which satisfy~\hbox{$\curl\vel \propto \vel$}.
This flow is time-independent, incompressible ($\detmetric=1$), and is
well-studied in the context of dynamo theory; we use the parameter
values~$A=5, B=C=2$, which are known to exhibit a vigorous
dynamo.\cite{Galloway1986,STF} (All the results presented here also apply to
the more common~$A=B=C=1$ flow, though the convergence rate of the constraints
is slower because that flow is less chaotic.)  Figure~\ref{fig:HumHmuABC522}
shows the convergence of the two terms of~\eqref{Hemconstr}
and~\eqref{Hmeconstr}; they are seen two track each other as time evolves.
Figure~\ref{fig:LogHumHmuABC522} is a logarithmic plot
of~$\helict_{\edir\mdir}$ and~$\helict_{\mdir\edir}$ as a function of time,
showing that the two quantities have converged to order~$10^{-7}$.  The
convergence is not linear: it exhibits large fluctuations, yet the decreasing
trend is evident.

Finally, no demonstration of a method would be complete without applying it to
the celebrated Lorenz model,\cite{Lorenz1963}
\begin{equation}
	\dot\x^1 = \sigma(\x^2 - \x^1),\qquad
	\dot\x^2 = r\,\x^1 - \x^2 - \x^1\,\x^3,\qquad
	\dot\x^3 = -b\,\x^3 + \x^1\,\x^2,
	\eqlabel{Lorenz}
\end{equation}
which is a compressible vector field with~\hbox{$\div\vel = -(\sigma+1+b)$}.
We use Lorenz's parameter values of~$\sigma=10$, $b=8/3$, $r=28$.  The initial
condition we used is~\hbox{$\lagrcv=(0,1,0)$}, which has a trajectory that
initially skirts an unstable fixed point and undergoes nearly-periodic
oscillations around it.  Thus, the convergence of the constraints is
momentarily delayed until the trajectory can explore the attractor, which is
seen to happen at~\hbox{$\time\simeq 20$} in
Figures~\ref{fig:HumHmuLorenz}--\ref{fig:HuuLorenz}.  Note that this does not
mean that if the trajectory comes near a fixed point again the convergence to
zero will be undone: it will pause momentarily, to resume after the trajectory
leaves the vicinity of the fixed point.

The off-diagonal terms~\eqref{Ricci3Doffdiag} involve the same quantities as
the diagonal ones, so they do not immediately yield any new constraints.
However, they may be useful in trying to establish a detailed curvature
balance for the Ricci tensor (see below).

It should be noted in closing this section that for the three-dimensional case
we have not made use of the fact that the curvature actually \emph{vanishes}.
We have simply assumed that its individual terms grow no faster
than~$\exp(2|\lyapexp_\sdir|\time)$.  In contrast to the two-dimensional case,
we have made no detailed assessment of the balance of curvature, a difficult
task because of the number of terms involved.  However, hints of this balance
can often be seen in the numerical results, such as
when~$\helict_{\edir\mdir}$ and~$\helict_{\mdir\edir}$ track each other
precisely in \figreftwo{LogHumHmuNF3_set01}{LogHumHmuLorenz}, or in
\figref{HuuABC522}, where~$\helict_{\mdir\mdir}$ and~$\helict_{\sdir\sdir}$
also track each other exactly.

\section{Discussion}
\seclabel{discussion}

Using the requirement that, in a flat space, the Riemannian curvature must
vanish in any coordinate system, we have derived the
constraints~\eqref{sdir2dconstraintsimp2} in two dimensions, a generalization
of the work of Tang and Boozer\cite{Tang1996} to compressible flows, with a
more detailed examination of the underlying assumptions.  We have also
verified the constraint for flows, whereas previous work had focused on
area-preserving maps.  Equation~\eqref{sdir2dconstraintsimp2} is a constraint
relating the stable direction,~$\sdiruv$, and the finite-time Lyapunov
exponent,~$\lyapexp_\sdir$; it is satisfied to exponential accuracy in time in
a chaotic flow.

With the same assumptions as in two dimensions, we have derived the same
constraint in three dimensions, Eq.~\eqref{sdir2dconstraintsimp2b}, verifying
the conjecture of Tang and Boozer\cite{Tang1999b} inspired by numerical
results on volume-preserving maps.  In three dimensions, it is a consequence
of two separate constraints, Eqs.~\eqref{Hemconstr} and~\eqref{Hmeconstr}.
There is also a third constraint, Eq.~\eqref{Heeconstr}, involving only the
unstable direction,~$\ediruv$, and no finite-time Lyapunov exponents.  All
have been shown numerically to hold with exponential accuracy in time for
$ABC$ flow and the Lorenz model, and we have verified the constraint for
several other flows and maps, parameter values, and initial conditions.  Two
of the three-dimensional constraints are satisfied trivially in two
dimensions, and the third is just the two-dimensional
constraint~\eqref{sdir2dconstraintsimp2}, showing the consistency of the
results.

The constraint~\eqref{sdir2dconstraintsimp2b} actually applies in any number
of dimensions, as long as the usual assumption holds: all the terms in the
Ricci scalar~\eqref{Ricciscal} must be bounded from above
by~$\exp(2|\lyapexp_\sdir|\time)$.  There is then
a~$\covder_i[\sdir^i\,\covder_j\sdir^j]$ term in the Ricci scalar that yields
the constraint~\eqref{sdir2dconstraintsimp2b}.  The higher-dimensional
generalization of Eqs.~\eqref{Hemconstr}, \eqref{Hmeconstr},
and~\eqref{Heeconstr} involve Lie brackets instead of curls, and a detailed
analysis of their form remains to be done.

Though the constraints are interesting from the mathematical standpoint, it is
natural to ask the following: what physical impact do the constraints have?
Previous work has focussed on the constraint~\eqref{sdir2dconstraintsimp2b} in
two\cite{Tang1996,Tang1999a} and three\cite{Tang1999b,Tang2000} dimensions.
We review this work and discuss the consequences of the new, three-dimensional
constraints.

In Ref.~\onlinecite{Tang1996}, it was found that the advection-diffusion
equation in Lagrangian coordinates is a diffusion equation with an anisotropic
diffusion tensor~\hbox{$\Diff^{ij} = \Diff\,\metric^{ij}$}, where~$\Diff$ is
the scalar molecular diffusivity and~$\metric^{ij}$ is the inverse of the
metric tensor given by Eq.~\eqref{metricdef}.  The dominant direction of
diffusion is along the~$\sdiruv$ line,
with~\hbox{$\sdiru_i\,\Diff^{ij}\,\sdiru_j =
\Diff\exp(-2\lyapexp_\sdir\,\time)$}, which grows exponentially
because~\hbox{$\lyapexp_\sdir < 0$}.  In contrast, the diffusion along
the~$\ediruv$ line is exponentially damped at a
rate~\hbox{$\exp(-2\lyapexp_\edir\,\time)$}.  Thus, after a short time
essentially all the diffusion occurs along the~$\sdiruv$ line.  This
counterintuitive result is a consequence of the diffusion equation smoothing
out strong gradients, which are being created along the~$\sdiruv$ line because
of the \emph{contraction} of fluid elements along that direction.  Thus, it is
the stable direction and~$\lyapexp_\sdir$---and not~$\lyapexp_\edir$---that
characterize the chaotic enhancement to diffusion.  This is where the
constraint comes in: in Ref.~\onlinecite{Tang1996} is was shown that, in two
dimensions, Eq.~\eqref{sdir2dconstraintsimp2} can be written
\begin{equation*}
	\sdiruv\cdot\gradlc(\lyapexp_\sdir\time - \ln\|\kcurv\|)
		+ \sdiruv\cdot(\kcurv\times\curllc\kcurv)/\|\kcurv\|^2
	= 0,
\end{equation*}
where~\hbox{$\kcurv = (\sdiruv\cdot\grad)\sdiruv$} is the curvature of
the~$\sdiruv$ line (unrelated to the Riemannian curvature of the space).  This
implies that, along an~$\sdiruv$ line, the variations in~$\lyapexp_\sdir$ are
related to the variations in~$\|\kcurv\|$.  Indeed, this is confirmed by
calculations using maps in two\cite{Tang1996} and three\cite{Tang1999b}
dimensions: regions of the~$\sdiruv$ line with large curvature have locally
small Lyapunov exponents.  We have performed a similar calculation on the
oscillating convection rolls system to demonstrate this correlation:
Figure~\ref{fig:scurvlyap} (top) shows the magnitude of the curvature~$\kcurv$
as a function of the distance along the~$\sdiruv$ line labeled $A$ to $B$ in
\figref{oscrollsman}, parametrized by~$\tau$.  The bottom part of
\figref{scurvlyap} is a plot of the instantaneous Lyapunov
exponent~\hbox{$\lyapexp = |\lyapexp_\sdir|$} (the flow is incompressible, so
the two exponents differ only by a sign) as a function or~$\tau$.  Notice the
local minima in~$\lyapexp$ whenever the curvature is maximal.  These locally
small values of the Lyapunov exponent hinder the enhancement of diffusion in
those regions.  Such considerations were used to provide design criteria for a
chemical reactor in Ref.~\onlinecite{Tang1999a}.  Numerically, the correlation
was also verified for three-dimensional volume-preserving maps in
Ref.~\onlinecite{Tang1999b}.

The constraint~\eqref{sdir2dconstraintsimp2} has also been used in the study
of the dynamo problem,\cite{Tang2000} where a magnetic field is embedded in a
conducting chaotic flow.  Boozer\cite{Boozer1992} and Boozer and
Tang\cite{Tang2000} have derived criteria that determine the spatial
distribution of the magnetic field in terms of the magnitudes of the Lyapunov
exponents.

The constraint on the unstable direction~$\ediruv$, Eq.~\eqref{Heeconstr},
potentially modifies these results.  In Refs.~\onlinecite{Tang2000}
and~\onlinecite{Boozer1992}, an asymptotic expression for the magnitude of the
induced current in Lagrangian coordinates was derived,
\begin{equation}
	j^2 \sim \geigen_\edir^3\l\{
		(\mathbf{B}\cdot\ediruv)
		\,\ediruv\cdot\curllc\ediruv
	\r\}^2 + O(\geigen_\edir^2\,\geigen_\mdir),
	\eqlabel{j2}
\end{equation}
where~$\mathbf{B}$ is the magnetic field, and only the fastest-growing term,
proportional to~$\geigen_\edir^3$, was kept.  The assumption was that there
was nothing special about~$\ediruv\cdot\curllc\ediruv$, so it should be
roughly of order one.  However, we have found that
typically~\hbox{$\ediruv\cdot\curllc\ediruv \sim
\geigen_\edir^{-1}\,\geigen_\mdir \rightarrow 0$} (see
Figures~\ref{fig:LogHuuNF3_set02}, \ref{fig:LogHuuNF3_set01},
\ref{fig:HuuABC522}, and~\ref{fig:HuuLorenz}).  Thus, the first term in
Eq.~\eqref{j2} is actually of lower order than the neglected term: for a
nonideal plasma with resistivity~$\eta$, the power~$\eta\,j^2$ needed to
sustain the dynamo is potentially much smaller than previously thought.  The
constraint~\eqref{Heeconstr} also modifies the growth rate of other
quantities, such as the parallel current~$\mathbf{j}\cdot\mathbf{B}$.  A more
detailed analysis remains to be done, where the constraints~\eqref{Hemconstr}
and~\eqref{Hmeconstr} might be found useful.

\acknowledgments

The authors thank Diego del-Castillo-Negrete for helpful suggestions.  This
work was supported by the National Science Foundation and the Department of
Energy under a Partnership in Basic Plasma Science grant,
No.~DE-FG02-97ER54441.


\begin{thebibliography}{10}
\bibitem{truetensor}
This is true for any tensor, but only the Riemann tensor has the property that
  it vanishes for flat spaces.
\bibitem{orthobasis}
An orthonormal basis is generally not a coordinate basis, \ie, it does not
  correspond to vectors~$\{\pd/\pd{\x'}^i\}$ induced by some coordinate
  transformation~${\x'}^i=f^i(\xv)$.
\bibitem{Riccirotc}
The derivatives in the definition of~$\Riccirotc$, Eq.~\eqref{Riccirotcdef},
  must be covariant, but we shall not make direct use of that definition. A
  straightforward way to evaluate~$\Riccirotc$ is to write it as\cite{Wald} \[
  \Riccirotc_{\lambda\mu\nu} = \half\l\{ (\w_\mu)_i\,[\w_\lambda,\w_\nu]^i -
  (\w_\nu)_i\,[\w_\lambda,\w_\mu]^i + (\w_\lambda)_i\,[\w_\mu,\w_\nu]^i\r\}, \]
  where the Lie bracket in terms of ordinary partial derivatives
  is~\hbox{$[v,w]^i = v^k\,\pd w^i/\pd\xi^k - w^k\,\pd v^i/\pd\xi^k$}.
\end{thebibliography}

\clearpage

%
%

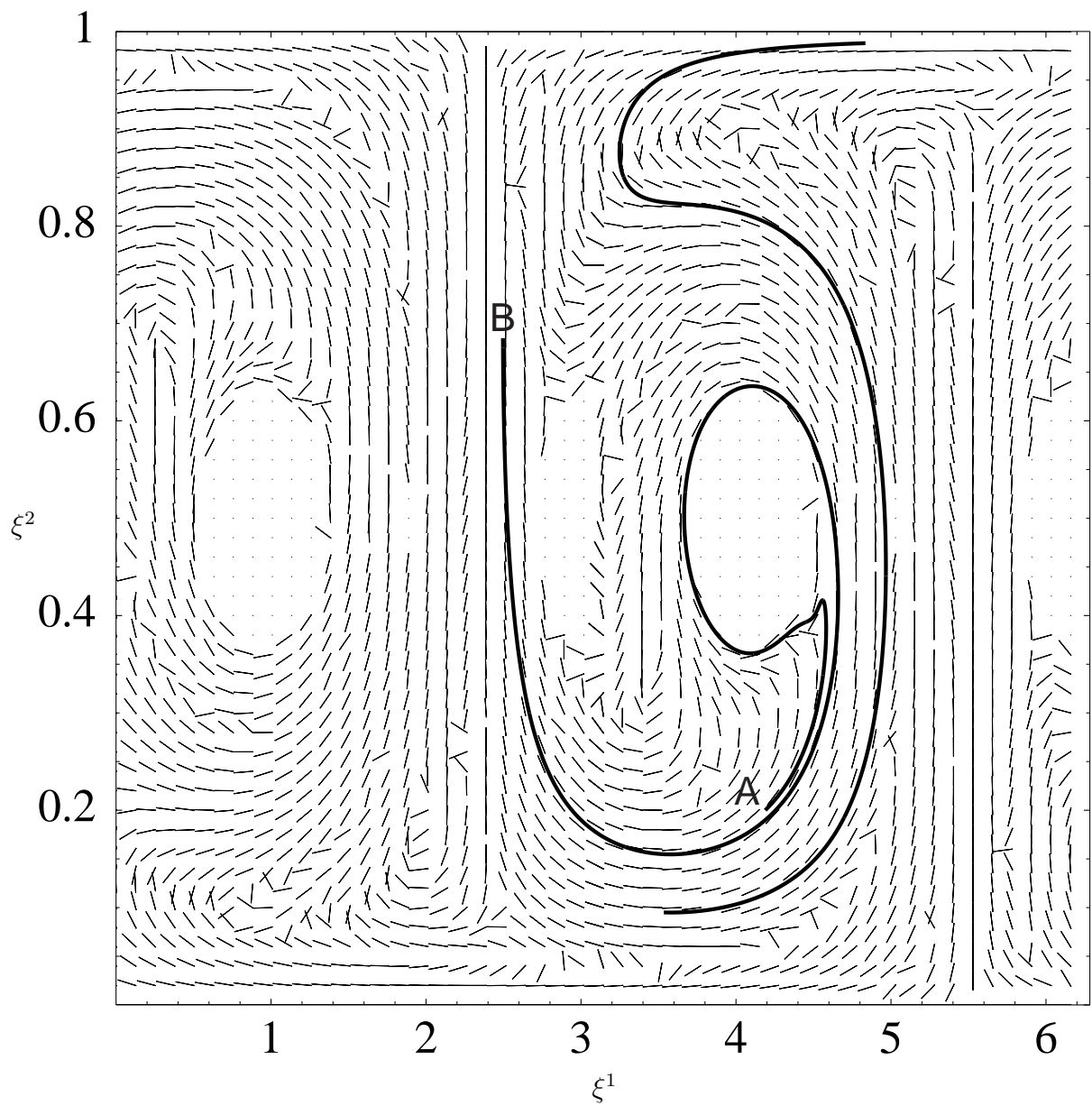
\begin{figure}
\caption{The field of characteristic directions $\sdiruv^\infty$ for
oscillating rolls, Eq.~\eqref{ocr}, with $A=k=\epsilon=\omega=1$.  Two typical
portions of the stable manifold are shown.}
\figlabel{oscrollsman}
\end{figure}

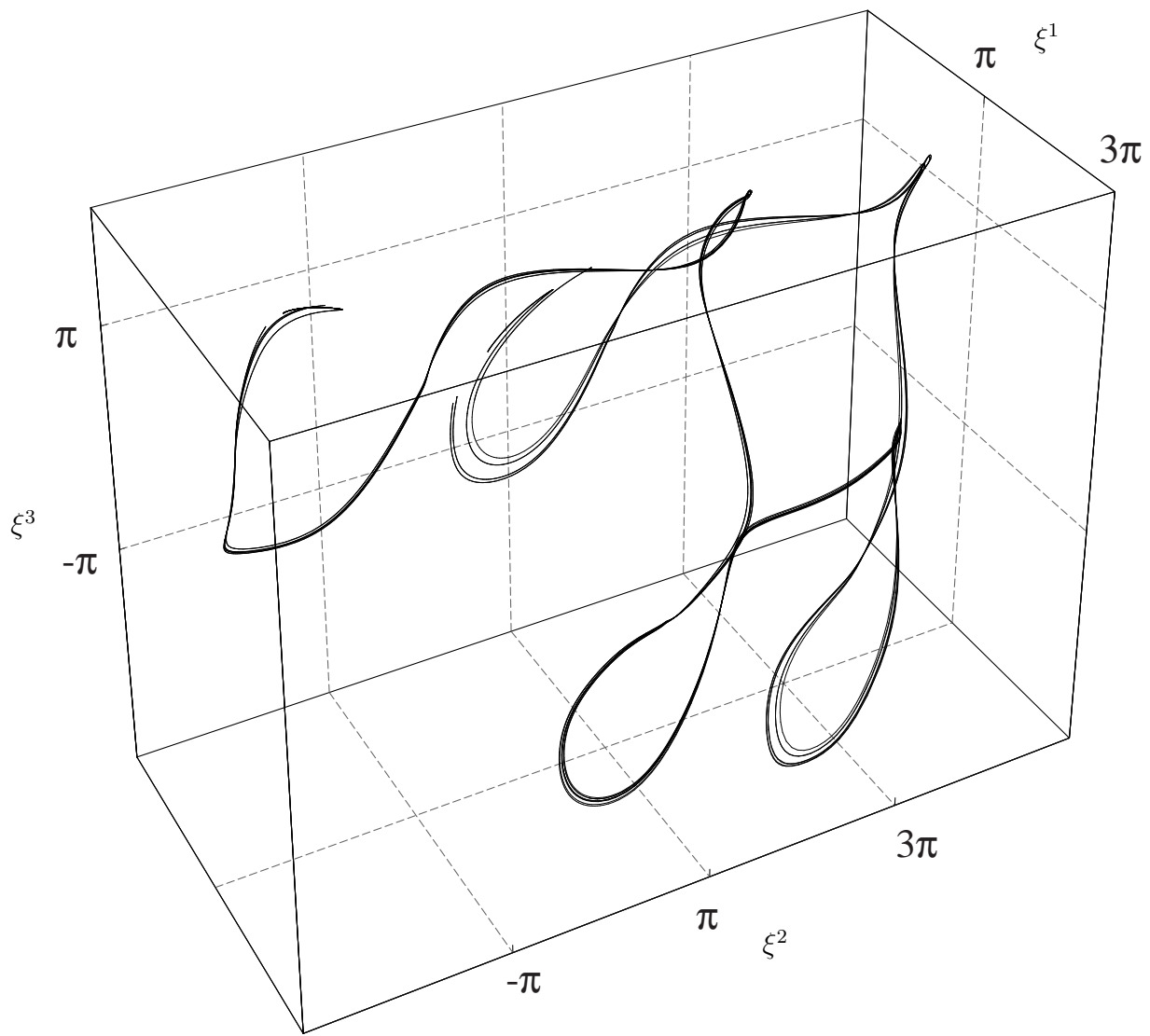
\begin{figure}
\caption{Typical portion of a stable manifold, or~$\sdiruv^\infty$-line, for
the~$ABC$ flow with~$A=5$, $B=C=2$.  The domain is~$2\pi$-periodic in all
three directions, as delimited by the dashed lines.}
\figlabel{ABC522sman}
\end{figure}

\psfrag{a}{$\tau$}

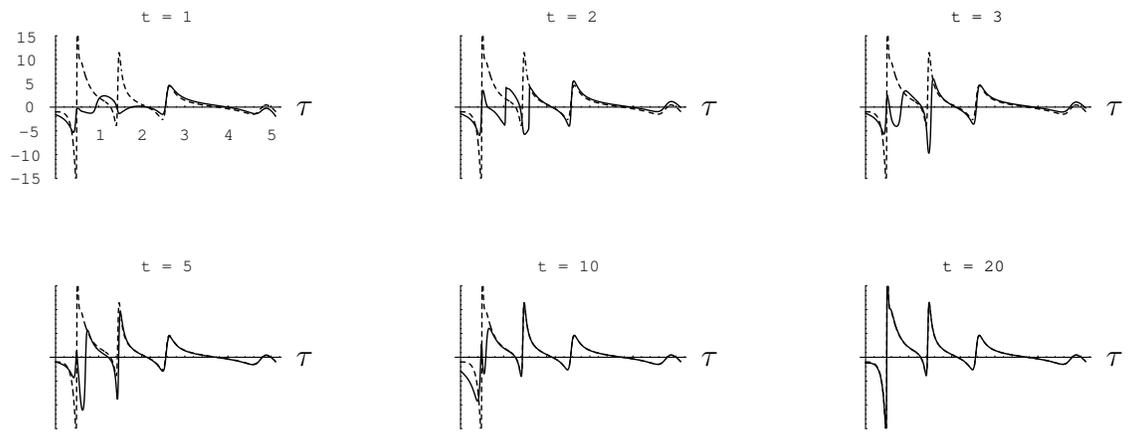
\begin{figure}
\caption{For the system of oscillating convection rolls, Eq.~\eqref{ocr}, the
convergence of the constraint~\hbox{$\divlc\sdiruv +
(\sdiruv\cdot\gradlc)\lyapexp_{\edir}\time \rightarrow 0$} is illustrated by
evaluating its two terms on an $\sdiruv^\infty$-line.  Here $\tau$ is the
distance along the $\sdiruv^\infty$-line from~$A$ to~$B$
in~\figref{oscrollsman}, the solid line is $-\divlc\sdiruv^\infty$, and the
dashed line is $(\sdiruv^\infty\cdot\gradlc)\lyapexp_{\edir}\time$.  Some
parts of the line converge more slowly because the finite-time Lyapunov
exponent is smaller in those regions (see \figref{scurvlyap}).}
\figlabel{twodslineconverg}
\end{figure}

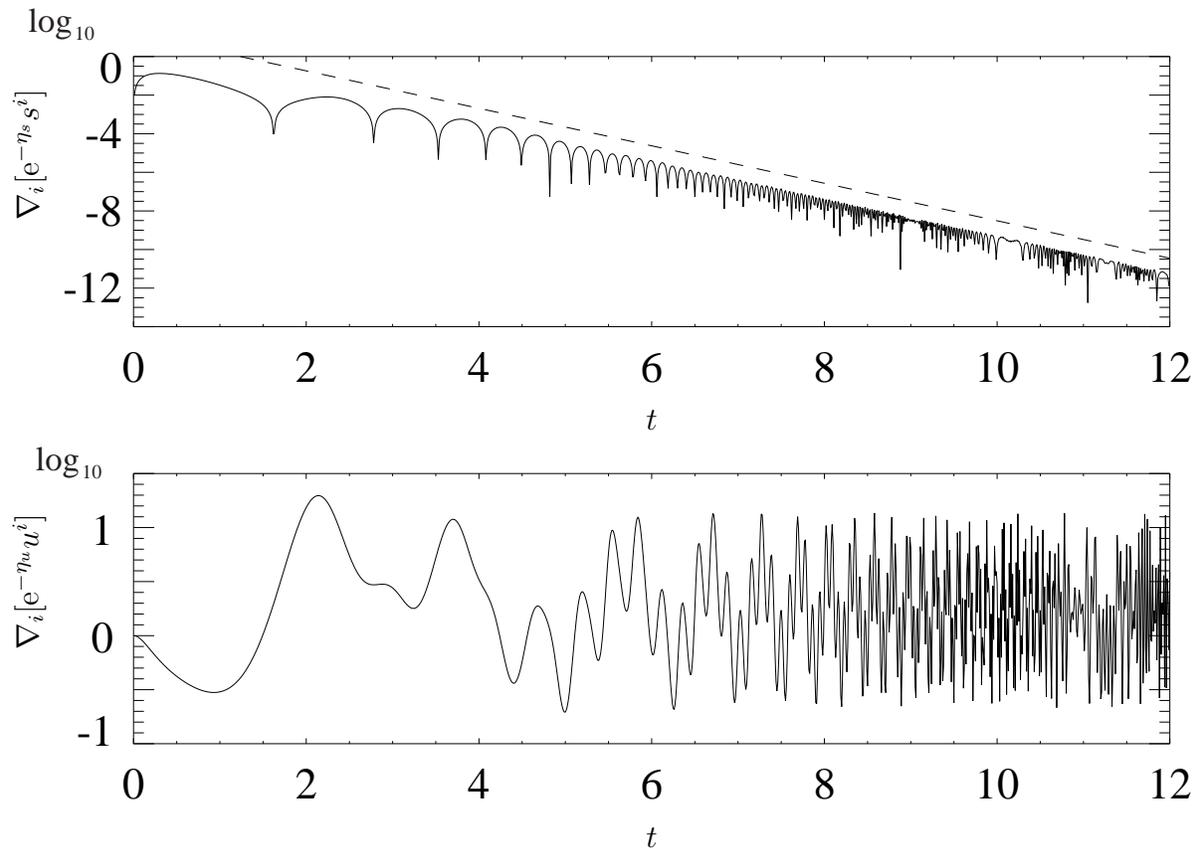
\begin{figure}
\caption{Top graph: Illustration of the exponential convergence
of~\hbox{$\divlc\sdiruv + \sdiruv\cdot\gradlc\lyapexp_\edir\,\time$}, for the
nonchaotic two-dimensional model given by Eq.~\eqref{nonchaotic2d}.  The
parameter values~$\mu=1$ and~$\nu=-1$ are chosen such that the trajectories
are exponentially diverging (the initial conditions
are~\hbox{$\lagrcv=(0.01,0.015)$}).  The dashed line denotes an exponential
with decay rate~\hbox{$-(\lyapexp_\edir + |\lyapexp_\sdir|) \simeq -(1.618 +
0.618) = -2.236$}.  Bottom graph: plot of~\hbox{$\divlc\ediruv +
\ediruv\cdot\gradlc\lyapexp_\sdir\,\time$}, showing that asymptotically it
remains of order one.  $\divlc\ediruv$ actually converges to a nonzero
constant; the oscillations are due
to~$\ediruv\cdot\gradlc\lyapexp_\sdir\,\time$.}
\figlabel{nonchaotic2dconverg}
\end{figure}

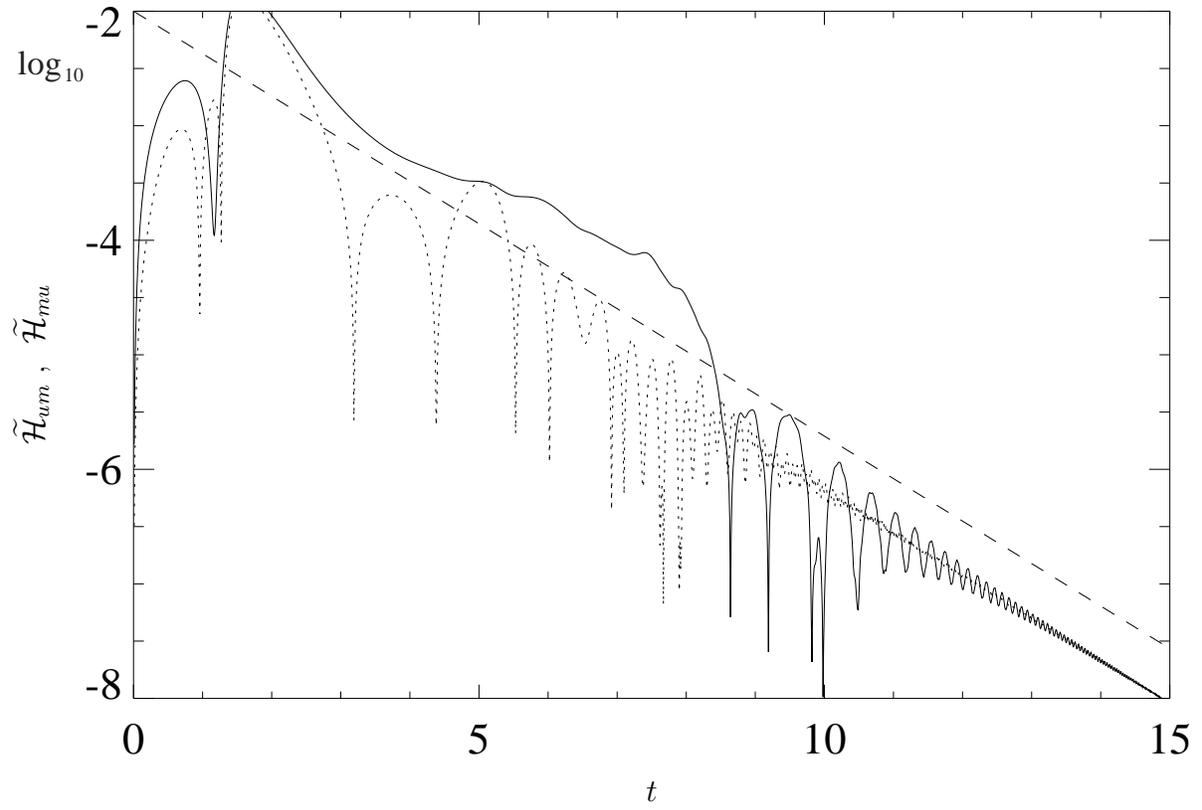
\begin{figure}
\caption{$\helict_{\edir\mdir}$ (solid line) and~$\helict_{\mdir\edir}$
(dotted line) for Eqs.~\eqref{nf3}, with parameter values~$\lambda = -0.1833$,
$\mu = 0.75$, $\nu = 0$ and initial
conditions~\hbox{$\lagrcv=(0.01,0.015,-0.018)$}.  For comparison, the dashed
line decreases as~$\exp(-2(\lyapexp_\edir - \lyapexp_\mdir)\time) \simeq
\exp(-2(0.698 - 0.271)\time) = \exp(-0.854\time)$.}
\figlabel{LogHumHmuNF3_set02}
\end{figure}

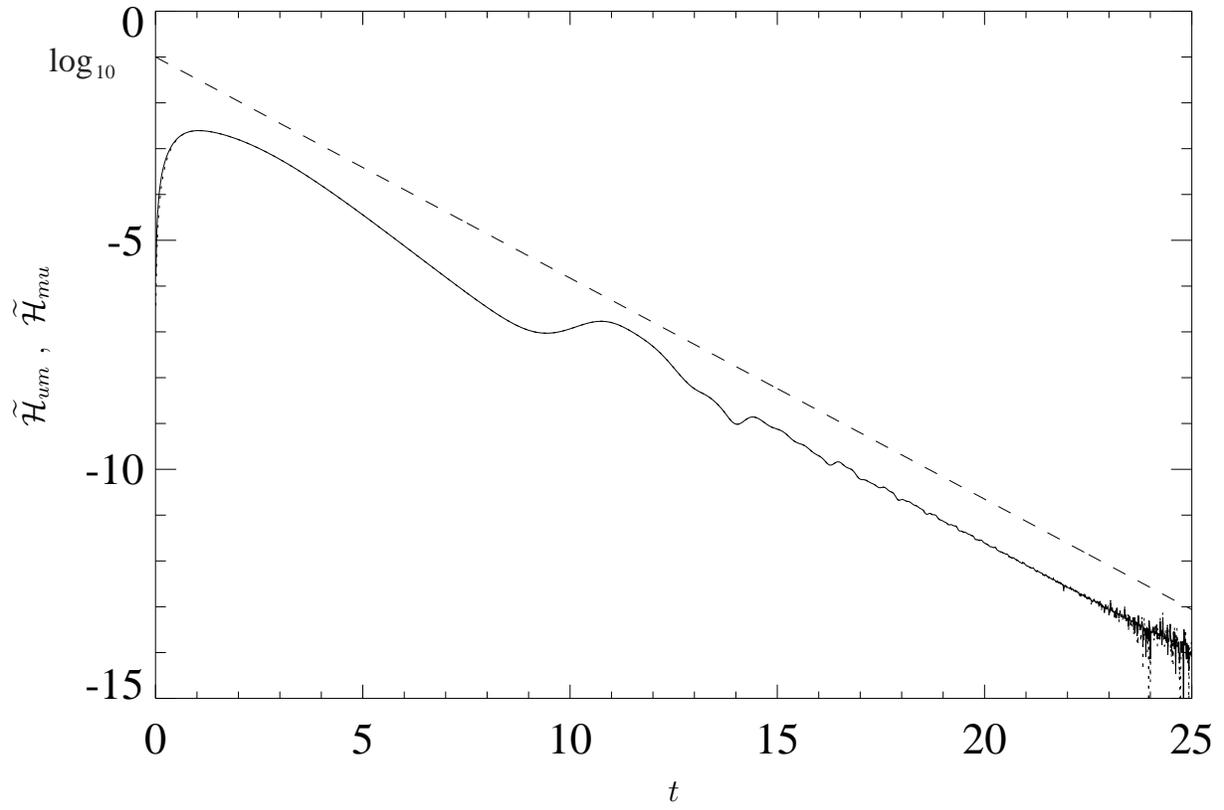
\begin{figure}
\caption{$\helict_{\edir\mdir}$ and~$\helict_{\mdir\edir}$ (respectively the
solid and dotted lines, which lie on top of each other) for Eqs.~\eqref{nf3},
with parameter values~$\lambda = 0.1833$, $\mu = 0.75$, $\nu = -2.6667$ and
initial conditions~\hbox{$\lagrcv=(0.01,0.015,-0.018)$}.  For comparison, the
dashed line decreases as~$\exp(-2(\lyapexp_\edir - \lyapexp_\mdir)\time)
\simeq \exp(-2(0.396 + 0.159)\time) = \exp(-1.11\time)$.  The noisiness
after~$\time\simeq 22$ is due to roundoff error.}
\figlabel{LogHumHmuNF3_set01}
\end{figure}

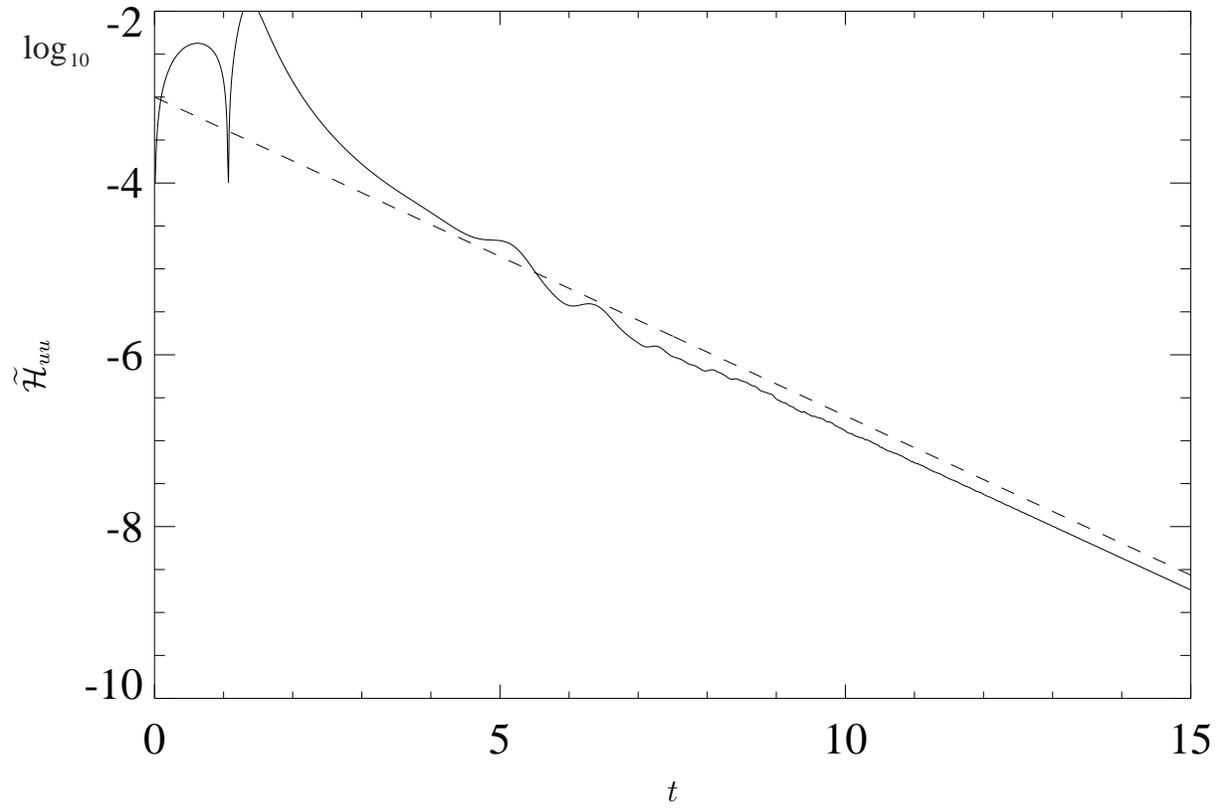
\begin{figure}
\caption{$\helict_{\edir\edir}$ (solid line) for Eqs.~\eqref{nf3}, with the
same parameter values and initial conditions as in \figref{LogHumHmuNF3_set02}.
The dashed line has the same slope as in \figref{LogHumHmuNF3_set02}.}
\figlabel{LogHuuNF3_set02}
\end{figure}

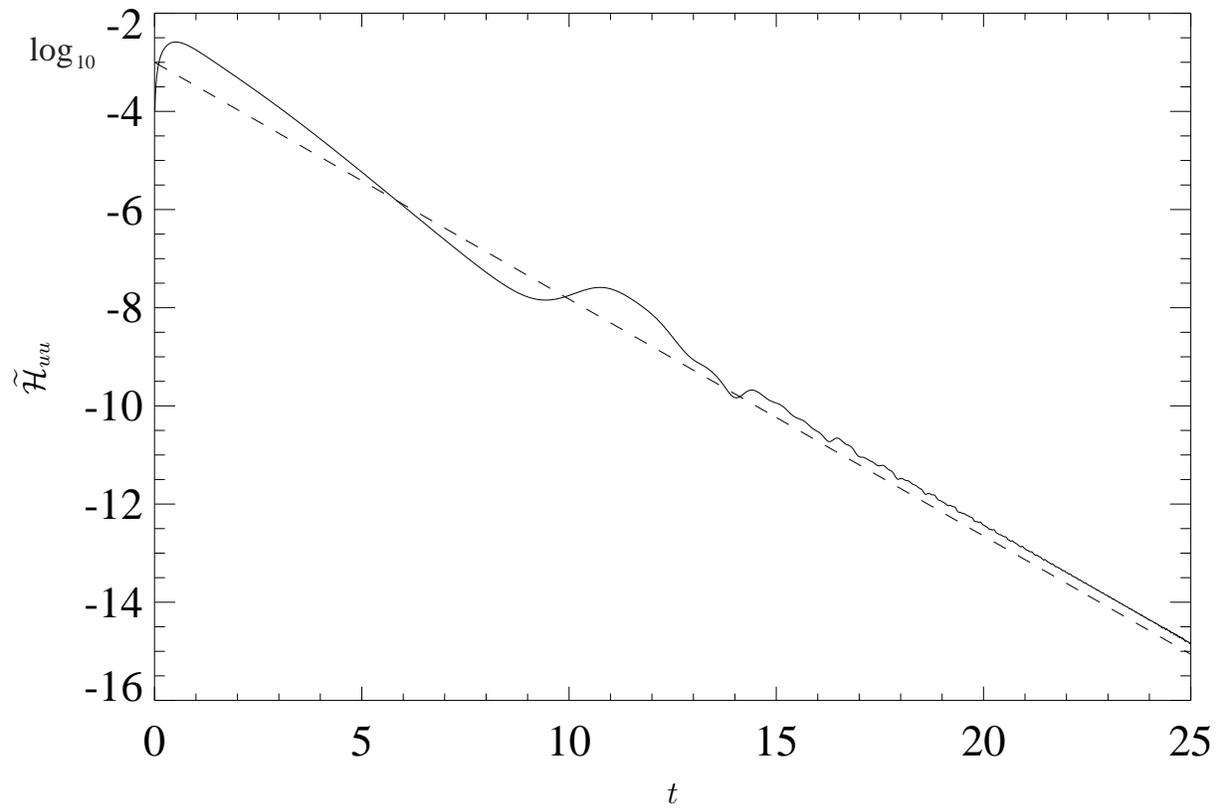
\begin{figure}
\caption{$\helict_{\edir\edir}$ (solid line) for Eqs.~\eqref{nf3}, with the
same parameter values and initial conditions as in \figref{LogHumHmuNF3_set01}.
The dashed line has the same slope as in \figref{LogHumHmuNF3_set01}.}
\figlabel{LogHuuNF3_set01}
\end{figure}

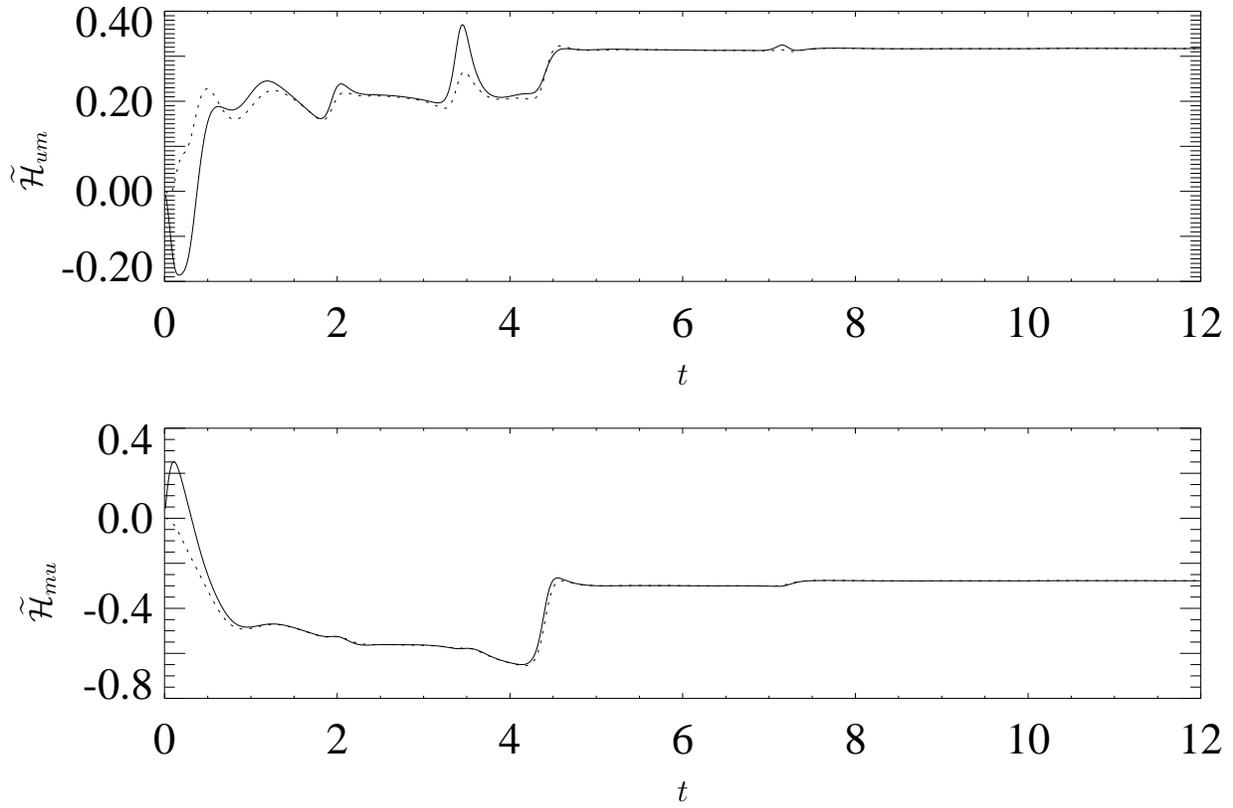
\begin{figure}
\caption{Top: Plot of~$\ediruv\cdot\curllc\mdiruv$ (solid line)
and~$\sdiruv\cdot\gradlc\lyapexp_\mdir\time$ (dotted line) versus time
for~$A=5$, $B=C=2$ flow [Eqs.~\eqref{ABCflow}], with initial
conditions~\hbox{$\lagrcv=(0.5,0.3,0.3232)$}.  The convergence of the two lines
verifies the constraint~\eqref{Hemconstr}.  Bottom: Plot
of~$\mdiruv\cdot\curllc\ediruv$ (solid line)
and~$-\sdiruv\cdot\gradlc\lyapexp_\edir\time$ (dotted line) versus time for
the same flow.  The convergence of the two lines verifies the
constraint~\eqref{Hmeconstr}.}
\figlabel{HumHmuABC522}
\end{figure}

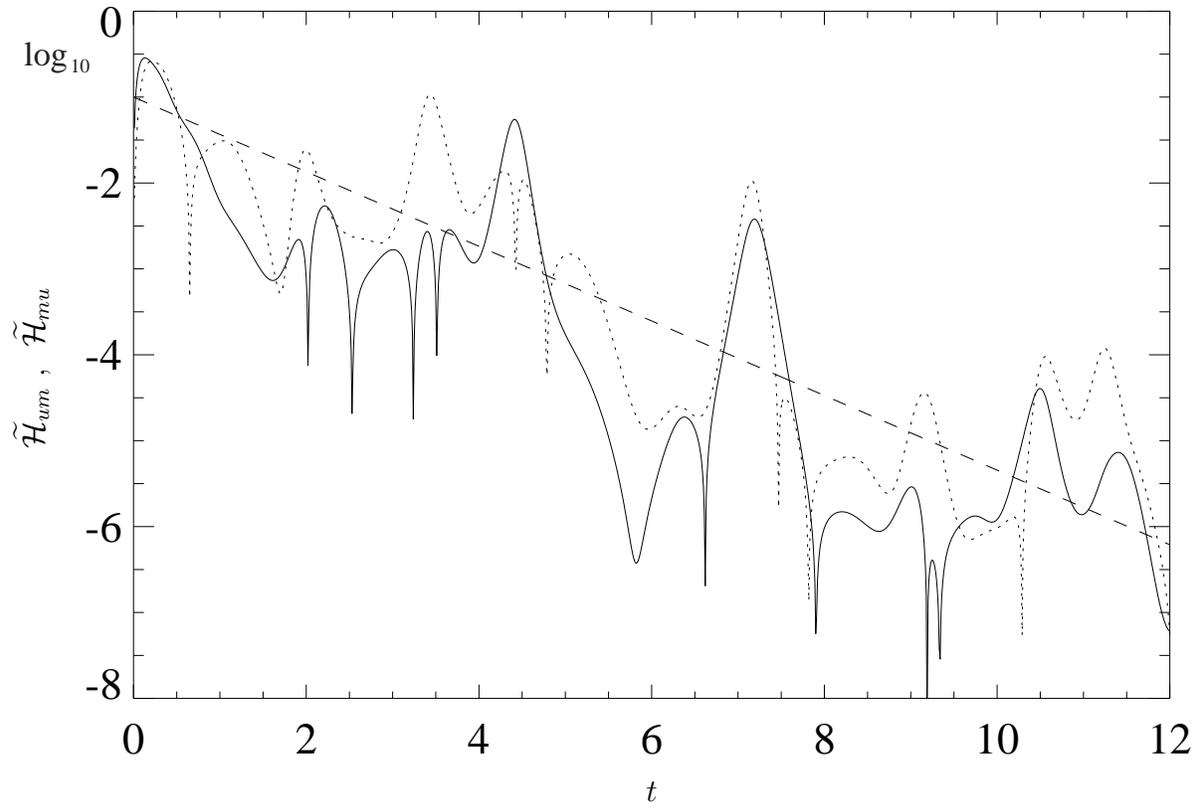
\begin{figure}
\caption{Log-linear (base 10) plot of~$\helict_{\edir\mdir}$ (solid line)
and~$\helict_{\mdir\edir}$ (dotted line) versus time for~$A=5$, $B=C=2$ flow
[Eqs.~\eqref{ABCflow}], with initial conditions as in \figref{HumHmuABC522}.
There are large fluctuations, but the trend is for both terms to go to zero;
they have reached values of~$10^{-7}$ at~$\time=12$.  Note that, from
\figref{HumHmuABC522}, the terms that cancel to give exponential convergence
to zero are roughly of order one, so they agree to about seven digits.  For
comparison, the dashed line decreases like~\hbox{$\exp(-2(\lyapexp_\edir -
\lyapexp_\mdir)\time) \simeq \exp(-2\time)$}.}
\figlabel{LogHumHmuABC522}
\end{figure}

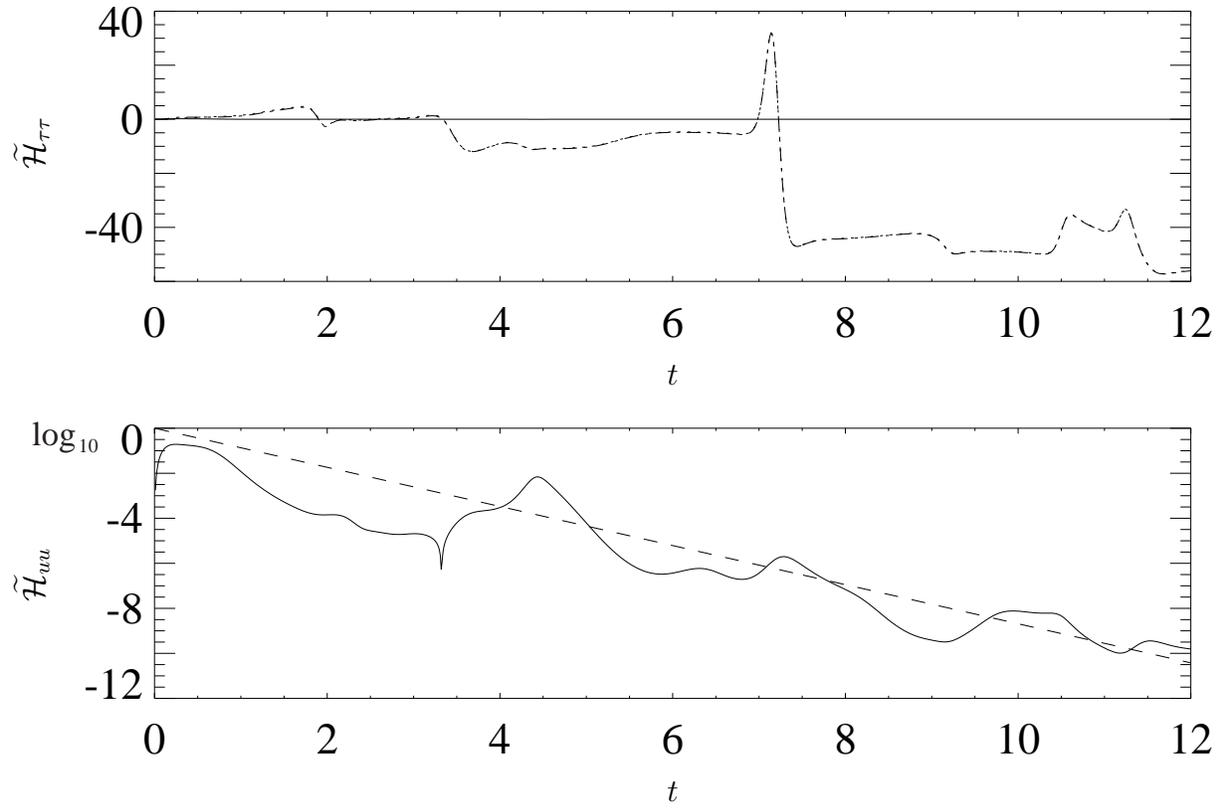
\begin{figure}
\caption{Top: Plot of~$\helict_{\edir\edir}$ (solid
line),~$\helict_{\mdir\mdir}$ (dashed line),~$\helict_{\sdir\sdir}$ (dotted
line, lying on top of~$\helict_{\mdir\mdir}$) versus time for~$A=5$, $B=C=2$
flow [Eqs.~\eqref{ABCflow}], with initial conditions as in
\figref{HumHmuABC522}.  For this particular system, with these parameters, we
have~$\helict_{\mdir\mdir}=\helict_{\sdir\sdir}$, though this does not appear
to be generic.  Bottom: Log-linear (base 10) plot
of~\hbox{$\helict_{\edir\edir} = \ediruv\cdot\curllc\ediruv$}, showing its
exponential convergence to zero---as predicted by the
constraint~\eqref{Heeconstr}.  For
comparison, the dashed line decreases like~\hbox{$\exp(-2(\lyapexp_\edir -
\lyapexp_\mdir)\time) \simeq \exp(-2\time)$}.}
\figlabel{HuuABC522}
\end{figure}

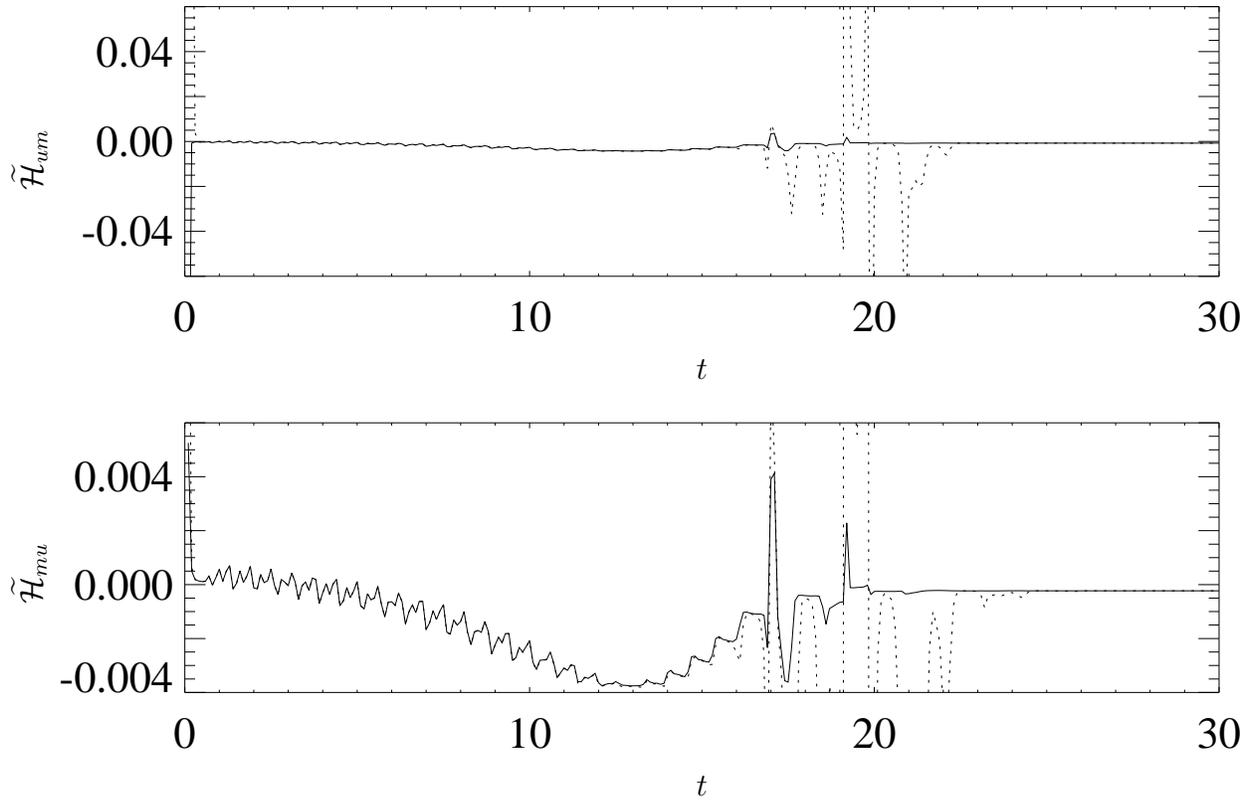
\begin{figure}
\caption{Top: Plot of~$\ediruv\cdot\curllc\mdiruv$ (solid line)
and~$\sdiruv\cdot\gradlc\lyapexp_\mdir\time$ (dotted line) versus time for the
Lorenz model [Eqs.~\eqref{Lorenz}], with parameter values~$\sigma=10$,
$b=8/3$, $r=28$, and initial conditions~\hbox{$\lagrcv=(0,1,0)$}.  Bottom:
Plot of~$\mdiruv\cdot\curllc\ediruv$ (solid line)
and~$-\sdiruv\cdot\gradlc\lyapexp_\edir\time$ (dotted line) versus time for
the same flow.  In both plots the two lines initially seem to converge but
then diverge abruptly before converging again at~$\time\simeq 21$.  This is
due to the trajectory approaching an unstable fixed point of the model and
undergoing nearly-periodic motion.  During that time the trajectory does not
``feel'' the chaos, and only after it leaves the vicinity of the fixed point
does convergence of~$\helict_{\edir\mdir}$ amd~$\helict_{\mdir\edir}$ resume.
(see \figref{LogHumHmuLorenz}.)}
\figlabel{HumHmuLorenz}
\end{figure}

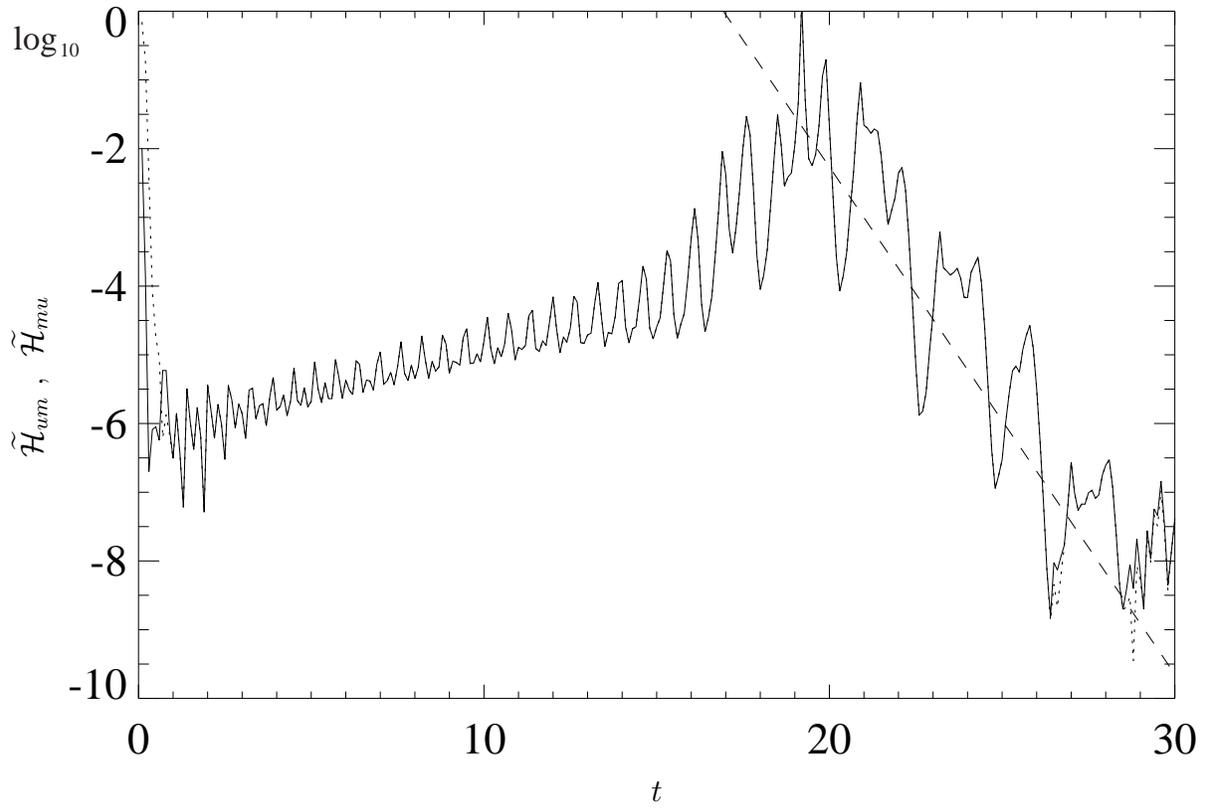
\begin{figure}
\caption{Log-linear (base 10) plot of~$\helict_{\edir\mdir}$ (solid line)
and~$\helict_{\mdir\edir}$ (dotted line) versus time for the Lorenz model with
the same parameter values as in \figref{HumHmuLorenz}.  As mentioned in
\figref{HumHmuLorenz}, the trajectory is initially nearly-periodic.  It has to
first get away from the unstable fixed point and explore the attractor before
the convergence really sets in, at~$\time\simeq 20$.}
\figlabel{LogHumHmuLorenz}
\end{figure}

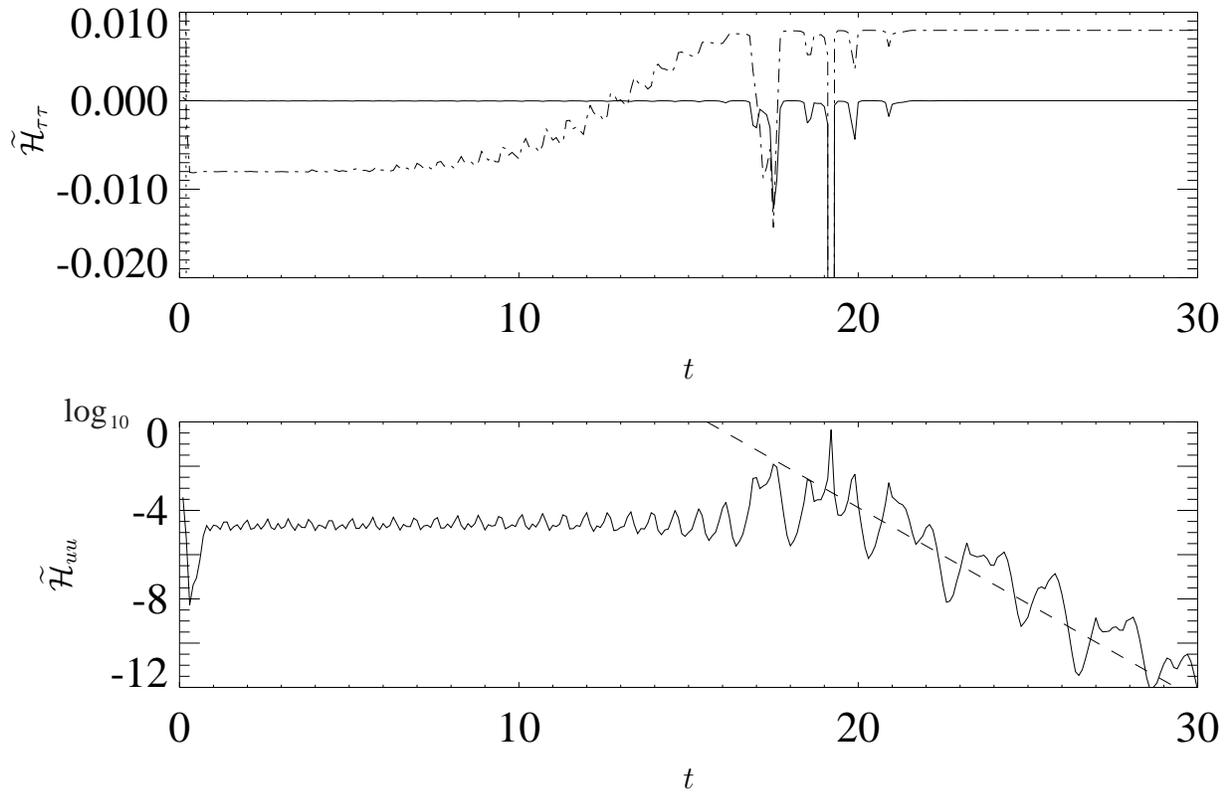
\begin{figure}
\caption{Top: Plot of~$\helict_{\edir\edir}$ (solid line)
and~$\helict_{\mdir\mdir}$ (dashed line) versus time for the Lorenz model with
the same parameter values as in \figref{HumHmuLorenz}.  For this particular
system, with these parameters, we
have~$\helict_{\mdir\mdir}=\helict_{\sdir\sdir}$, though this does not appear
to be generic.  Bottom: Log-linear (base 10) plot
of~\hbox{$\helict_{\edir\edir} = \ediruv\cdot\curllc\ediruv$}, showing its
exponential convergence to zero---as predicted by the
constraint~\eqref{Heeconstr}.  See the caption to
\figreftwo{HumHmuLorenz}{LogHumHmuLorenz} and the text for a discussion on the
transient oscillations before convergence sets in.}
\figlabel{HuuLorenz}
\end{figure}

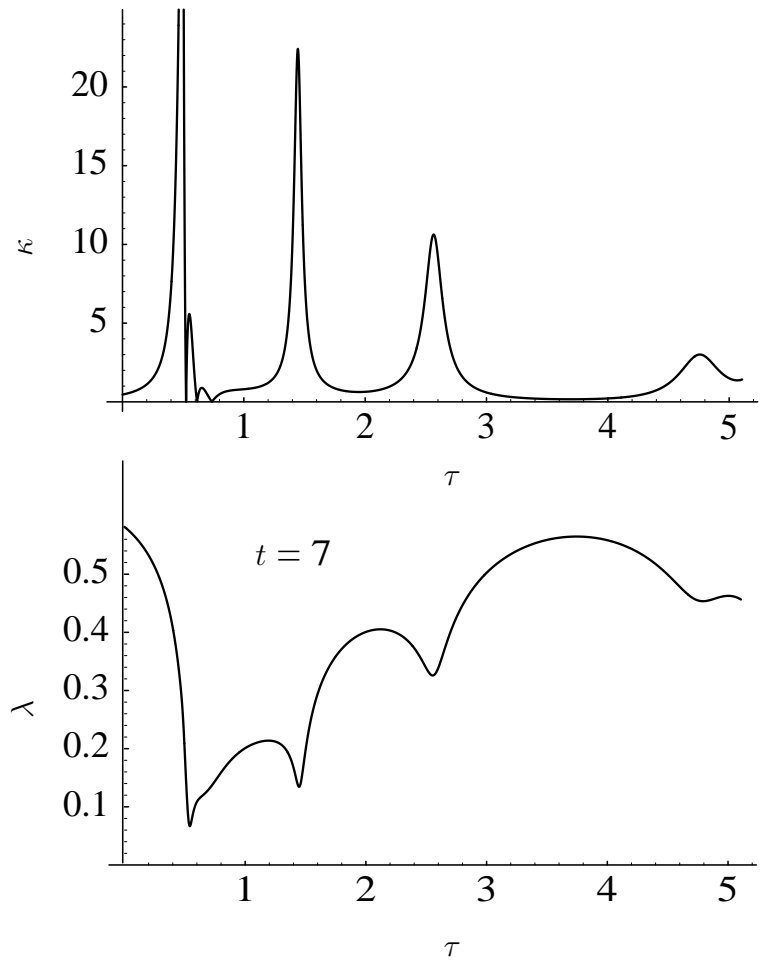
\begin{figure}
\caption{Top: For the system of oscillating convection rolls, a plot of the
magnitude of the curvature of the~$\sdiruv$ line,~$\kcurv =
(\sdiruv\cdot\gradlc)\sdiruv$, as a function of the distance~$\tau$ along
the~$\sdiruv$ line labeled $A$ to $B$ in \figref{oscrollsman} (with the same
parameter values).  Bottom: Plot of the finite-time Lyapunov
exponent~$\lyapexp$ at~$\time=7$ for the same system, as a function of~$\tau$.
Note the correlation between the locally small values of~$\lyapexp$ and the
magnitude of the curvature~$\kcurv$.}
\figlabel{scurvlyap}
\end{figure}

\clearpage

%
%

\newcommand{\figtopskip}{\vspace*{4em}}
\newcommand{\figlabskip}{\vspace{4em}}

\begin{figure}
\psfrag{x}{$\lagrc^1$}
\psfrag{y}{$\lagrc^2$}
\figtopskip
\centerline{\psfig{file=eps/convecrolls_sfield_sman.eps,width=\textwidth}}
\figlabskip\figref{oscrollsman}
\end{figure}

\begin{figure}
\psfrag{x}{$\lagrc^1$}
\psfrag{y}{$\lagrc^2$}
\psfrag{z}{$\lagrc^3$}
\figtopskip
\centerline{\psfig{file=eps/sman_abc522.eps,width=\textwidth}}
\figlabskip\figref{ABC522sman}
\end{figure}

\psfrag{a}{$\tau$}

\begin{figure}
\figtopskip
\centerline{\psfig{file=eps/converg.eps,width=\textwidth}}
\figlabskip\figref{twodslineconverg}
\end{figure}

\clearpage

\psfrag{u}{$\helict_{\edir\mdir}$}
\psfrag{m}{$\helict_{\mdir\edir}$}
\psfrag{h}{$\helict_{\edir\edir}$}
\psfrag{H}{$\helict_{\edir\mdir}$\ ,\ \ $\helict_{\mdir\edir}$}
\psfrag{M}{$\helict_{\tau\tau}$}
\psfrag{t}{$\time$}
\psfrag{k}{$\kcurvm$}
\psfrag{b}{$\lyapexp$}
\psfrag{d}{$\covder_i[\ee^{-\lyapexpt_\sdir}\sdir^i]$}
\psfrag{D}{$\covder_i[\ee^{-\lyapexpt_\edir}\edir^i]$}

\begin{figure}
\figtopskip
\centerline{\psfig{file=eps/Log_divs_divu_nonchaotic2d.eps,width=\textwidth}}
\figlabskip\figref{nonchaotic2dconverg}
\end{figure}

\begin{figure}
\figtopskip
\centerline{\psfig{file=eps/LogHumHmuNF3_set02.eps,width=\textwidth}}
\figlabskip\figref{LogHumHmuNF3_set02}
\end{figure}

\begin{figure}
\figtopskip
\centerline{\psfig{file=eps/LogHumHmuNF3_set01.eps,width=\textwidth}}
\figlabskip\figref{LogHumHmuNF3_set01}
\end{figure}

\begin{figure}
\figtopskip
\centerline{\psfig{file=eps/LogHuuNF3_set02.eps,width=\textwidth}}
\figlabskip\figref{LogHuuNF3_set02}
\end{figure}

\begin{figure}
\figtopskip
\centerline{\psfig{file=eps/LogHuuNF3_set01.eps,width=\textwidth}}
\figlabskip\figref{LogHuuNF3_set01}
\end{figure}

\begin{figure}
\figtopskip
\centerline{\psfig{file=eps/HumHmuABC522.eps,width=\textwidth}}
\figlabskip\figref{HumHmuABC522}
\end{figure}

\begin{figure}
\figtopskip
\centerline{\psfig{file=eps/LogHumHmuABC522.eps,width=\textwidth}}
\figlabskip\figref{LogHumHmuABC522}
\end{figure}

\begin{figure}
\figtopskip
\centerline{\psfig{file=eps/HuuABC522.eps,width=\textwidth}}
\figlabskip\figref{HuuABC522}
\end{figure}

\begin{figure}
\figtopskip
\centerline{\psfig{file=eps/HumHmuLorenz.eps,width=\textwidth}}
\figlabskip\figref{HumHmuLorenz}
\end{figure}

\begin{figure}
\figtopskip
\centerline{\psfig{file=eps/LogHumHmuLorenz.eps,width=\textwidth}}
\figlabskip\figref{LogHumHmuLorenz}
\end{figure}

\begin{figure}
\figtopskip
\centerline{\psfig{file=eps/HuuLorenz.eps,width=\textwidth}}
\figlabskip\figref{HuuLorenz}
\end{figure}

\begin{figure}
\figtopskip
\centerline{\psfig{file=eps/scurvlyap.eps,width=4in}}
\figlabskip\figref{scurvlyap}
\end{figure}

\end{document}